\documentclass[aps, prapplied, final, reprint, floatfix, notitlepage, longbibliography]{revtex4-2}
\usepackage[T1]{fontenc}
\usepackage[utf8]{inputenc}
\usepackage{bm, eucal, amssymb, amsmath, graphicx, color}
\usepackage{hyperref}
\hypersetup{colorlinks=true, citecolor=blue, linkcolor=magenta}

\newcommand{\vect}[1]{\mathbf{#1}}
\newcommand{\mat}[1]{\mathsf{#1}}
\newcommand{\ten}[1]{\mathsf{#1}} 
\newcommand{\Cten}{\ten{C}}
\newcommand{\Emat}{\mat{E}}
\newcommand{\Gmat}{\mathsf{\Gamma}}
\newcommand{\Amat}{\mat{A}}
\newcommand{\Kmat}{\mat{K}}
\newcommand{\Mmat}{\mat{M}}
\newcommand{\Qmat}{\mat{Q}}
\newcommand{\Vspace}{\mathcal V}
\newcommand{\Cphys}{\mathcal C_{\rm phys}}
\newcommand{\Melas}{\mathcal M}
\newcommand{\Oiso}{\mathcal O}
\newcommand{\Sym}{\mathbb S}
\newcommand{\RR}{Rayleigh-Ritz}

\graphicspath{{./figures/}}

\begin{document}

\title{
Physics-informed learning for the inverse problem in resonant ultrasound spectroscopy
}

\author{Alejandro Cubillos Muñoz}
\author{Manuela Rivas}
\author{Julián Rincón}
\affiliation{Department of Physics, Universidad de los Andes, Bogotá, D.C. 111711, Colombia}

\date{\today}

\begin{abstract}
Inferring elastic constants from resonant ultrasound spectra is a nonlinear and typically overdetermined inverse problem based on finite spectral data. We formulate the \RR{} inverse problem as a constrained inverse-isospectral problem on the set of physically admissible elasticity tensors. This induces effective low-dimensional variables for the inverse map on the admissible elasticity manifold: length and elastic scales, aspect-ratio coordinates, scale-free spectral features, and stability-respecting elastic ratios. We use these variables to construct a physics-informed learning pipeline in which a regression model acts only on reduced spectral and geometric features, while scale recovery and final elastic-constant reconstruction are imposed analytically. For the full cubic benchmark, the reconstructed constants have MAE values of $20.37(35.15)$, $24.30(41.33)$, and $2.13(3.66)~\mathrm{GPa}$ for $C_{11}$, $C_{12}$, and $C_{44}$. In the fixed-geometry benchmark, the corresponding cubic MAPE values are $4.14(3.87)\%$, $8.31(8.50)\%$, and $2.44(2.86)\%$, while the isotropic values are $4.0(3.6)\%$ and $0.4(0.3)\%$ for the bulk and shear moduli. The inverse problem then becomes a constrained regression problem in variables adapted to the geometry, scaling, crystal symmetry, and thermodynamic stability of Hookean elasticity.
\end{abstract}

\maketitle


\section{Introduction}

Elastic constants describe the linear mechanical response of a solid and encode information about its microscopic structure and thermodynamic behavior. They are related to the curvature of the free energy and are therefore useful for mechanical characterization, phase transitions, anharmonicity, and couplings between elastic and other excitations~\cite{Maynard1996RUS, Landau1986Elasticity, Sokolnikoff1956, Langhaar2016, Leisure2017Ultrasonic}. Accurate elastic constants are thus central to materials characterization and low-energy mechanical response.

Resonant ultrasound spectroscopy (RUS) infers elastic constants from the vibration frequencies of a solid~\cite{Migliori1993RUS, Migliori1997RUSBook, Maynard1996RUS, Leisure1997RUS, Schwarz2000RUS, Balakirev2019RUSToolbox, Geimer2026}. Acoustic and ultrasonic resonances are also used as condensed-matter probes, for example in studies of valence fluctuations, hidden order, and superconductivity~\cite{Maynard2001AcousticalAnalogs, Maynard2024Acoustics, Ramshaw2015Plutonium, Ghosh2020URu2Si2, Theuss2024UTe2}. In RUS, resonance peaks are extracted from the mechanical response over a frequency window. The forward problem computes these resonances from the elasticity tensor and sample data; after a \RR{} approximation, it becomes a finite-dimensional generalized eigenvalue problem in the class of variational normal-mode problems familiar from classical mechanics~\cite{Thorne2017ModernClassical}. The inverse problem is to infer elastic constants from a finite set of measured resonances and sample data.

This inverse step is not a direct inversion of the forward calculation. Resonance frequencies depend nonlinearly on the elastic constants because the mode shapes vary with material parameters. The measured data are finite, may contain uncertain peaks, and do not include mode shapes. At the same time, crystal symmetry fixes the number of independent constants, and thermodynamic stability restricts the admissible elasticity tensors. RUS inversion is therefore better viewed as a constrained inverse-isospectral problem: the spectrum imposes constraints, but admissible solutions must come from physically stable Hookean elasticity.

Machine-learning models can approximate nonlinear inverse maps, and recent work has used data-driven approaches for RUS and related elastic-inversion problems~\cite{Fukuda2023DLRUS, Yang2022, Liu2023, Shi2025InverseRUS, CubillosMunoz2025}. However, direct regression from raw dimensions, density, and resonance frequencies leaves known structure for the model to rediscover. The resonances are correlated because they depend on the same elastic constants. Absolute size, aspect ratio, elastic scale, and stability constraints enter the problem in distinct ways; treating them as raw correlations obscures the geometry of the inverse problem and enlarges the learning task.

The approach developed here is to expose this structure before learning. We formulate RUS inversion in the \RR{} representation as a constrained inverse-isospectral problem and introduce variables that separate size from aspect ratio, spectral scale from spectral shape, and elastic magnitude from stability-respecting elastic ratios. In these coordinates, the regression model is not asked to learn the homogeneity of the forward problem, the stability inequalities, or the final reconstruction formulas; it learns the reduced map between spectral-shape variables and bounded elastic-ratio coordinates.

Although the formalism is developed for RUS, the broader strategy of using physics-adapted variables before learning may be useful in other ultrasonic elastic-characterization methods. Impulse-excitation techniques infer elastic properties from resonance frequencies, while pulse-echo methods infer them from wave velocities~\cite{Popov2020IET, Grimes2011UPE}. In RUS, each normal mode generally depends on several elastic constants and the sample geometry. RUS provides a controlled setting in which this strategy can be formulated as a constrained inverse-isospectral problem.

The paper is organized as follows. Section~\ref{sec:rus_inverse_spectral} develops the forward and inverse problems and the constrained isospectral description. Section~\ref{sec:physics_informed_reduction} presents the physics-informed reduction and learning pipeline. Section~\ref{sec:results_discussion} gives the benchmarks and physical interpretation, and Sec.~\ref{sec:conclusions_outlook} the conclusions and outlook. Appendices~\ref{app:rayleigh_ritz_geometry}~and~\ref{app:supplementary_data_benchmarks} provide \RR{} details and further results.

\section{Constrained inverse spectral formulation of RUS}
\label{sec:rus_inverse_spectral}

\subsection{Spring-mass analogs for RUS problems}

The physical and mathematical structure of RUS can be motivated from basic spring-mass oscillators~\cite{KleppnerKolenkow2014, RossingFletcher2004}. These models isolate the finite-dimensional spectral logic that appears after a \RR{} approximation: inertial, geometric, and stiffness data determine resonances, while measured resonances constrain the inverse reconstruction. A driven oscillator illustrates the forward response, and a two-mass system shows how the inverse problem depends on physically admissible parameters.

The simplest forward model with the same logic as RUS is the driven damped oscillator, $m\ddot x+\gamma\dot x+kx=F_0\cos\Omega t$~\cite{KleppnerKolenkow2014, RossingFletcher2004}. The spring constant $k$ plays the role of the stiffness, while the driving force is the analog of the transducer excitation. For a steady-state response $x(t)=\operatorname{Re}[\tilde x(\Omega)e^{i\Omega t}]$, one has $\tilde x(\Omega)=\chi(\Omega)F_0$, with $\chi(\Omega)=(k-m\Omega^2+i\gamma\Omega)^{-1}$. Thus, if $m$, $\gamma$, and $k$ are known, the frequency response is predicted. Although this response is not globally Lorentzian, its absorption part, $-\operatorname{Im}\chi(\Omega)$, becomes locally Lorentzian near an isolated weakly damped resonance. Real RUS spectra contain many modes, mode-dependent damping, backgrounds, and possible peak overlap, but the logic is the same: known sample and stiffness parameters determine the resonant response.

The same idea extends to finite oscillatory systems, where normal modes, $\vect q\in\mathbb R^n$, satisfy~\cite{RossingFletcher2004}
\begin{equation}
    \Kmat\vect q=\omega^2\Mmat\vect q,
    \label{eq:discrete_generalized_eigenproblem}
\end{equation}
where $\Mmat,\Kmat\in\Sym^n$ are the mass and stiffness matrices, and $\Sym^n$ denotes the space of real symmetric $n\times n$ matrices. The solution to the forward problem can be encoded in a map $(\Mmat,\Kmat)\mapsto\{\omega_j\}$. This generalized eigenvalue problem has the same algebraic form as the \RR{} approximation used in RUS~\eqref{eq:rus_generalized_eigenproblem_main}: $\Mmat$ and $\Kmat$ are replaced by the Gram matrix $\Emat(\rho)$ and the stiffness matrix $\Gmat(\Cten)$, while the spring constants $\{k_j\}$ are replaced by the elasticity tensor $\Cten$. The sample geometry and crystal symmetry determine which matrices can arise, and crystal symmetry fixes the independent elastic constants.

A linear model of longitudinal oscillations illustrates the inverse problem. Consider two identical masses $m$ connected by three springs: $k_1$ and $k_3$ to fixed walls, and $k_2$ between masses. The walls remove the rigid translational zero mode. The stiffness and mass matrices are
\begin{equation}
    \Kmat=\begin{pmatrix}
    k_1+k_2 & -k_2\\
    -k_2 & k_2+k_3
    \end{pmatrix},
    \quad
    \Mmat=m\mat{I}_2 .
    \label{eq:two_mass_stiffness_matrix}
\end{equation}
The normal modes obey~\eqref{eq:discrete_generalized_eigenproblem}. Defining $\nu_\pm=m\omega_\pm^2$, the two eigenvalues are
\begin{equation}
    \nu_\pm
    =
    \frac{k_1+2k_2+k_3}{2}
    \pm
    \frac{1}{2}
    \sqrt{(k_1-k_3)^2+4k_2^2}.
    \label{eq:two_mass_general_eigenvalues}
\end{equation}
Thus the inverse problem is not determined by the number of measured frequencies alone; it also depends on the admissible stiffness structure and on whether that structure makes recovery effectively linear.

If all springs are constrained to be equal, $k_1=k_2=k_3=k$, then $\nu_-=k$ and $\nu_+=3k$. There is one unknown parameter but two measured eigenvalues, so the inverse problem is overdetermined: exact data must satisfy $\nu_+=3\nu_-$. With noisy data, the task becomes estimation of the single scale $k$ from a spectral mismatch.

If the wall springs are equal but the central spring is different, $k_1=k_3=k_w$ and $k_2=k_c$, then $\nu_-=k_w$ and $\nu_+=k_w+2k_c$. There are two unknowns and two measured eigenvalues, so the inverse problem is determined, with $k_w=\nu_-$ and $k_c=(\nu_+-\nu_-)/2$.

If all three $\{k_j\}$ are independent, then the inverse problem has three unknowns but only two eigenvalues. The spectrum fixes only the trace, $\nu_++\nu_-=k_1+2k_2+k_3$, and the determinant, $\nu_+\nu_-=k_1k_2+k_1k_3+k_2k_3$, invariants. 
Thus the inverse problem is underdetermined unless additional information or constraints are imposed. This example also separates parameter counting from nonlinearity. $\Kmat$ is linear in $\{k_j\}$, but $\omega^2$ are not in general. When $\{k_j\}$ are independent, the trace constraint is linear, whereas the determinant constraint is quadratic. Thus the inverse spectral relation is nonlinear even in this elementary system. By contrast, in the constrained cases above the imposed structure makes the recovery effectively linear: when all springs are equal, the spectrum depends on a single scale parameter, and when $k_1=k_3$ the symmetry fixes the normal-mode combinations, giving the linear relations $\nu_-=k_w$ and $\nu_+=k_w+2k_c$.

This example can be read in the scale-reduced language used below for RUS. Since the masses are known, a common rescaling of all spring constants rescales the squared frequencies; after removing this scale, the remaining spectral information constrains stiffness ratios. In the wall-symmetric case, the two unknowns are one stiffness scale and one ratio. The scale-free quantities
\begin{equation}
    \frac{\nu_-}{\nu_+},
    \qquad
    \frac{\nu_+-\nu_-}{\nu_+}
    \label{eq:two_mass_scale_free_features}
\end{equation}
are invariant under common rescaling of $\{k_j\}$ and are the two-mode analogs of the ratio~\eqref{eq:xi_definition} and normalized-gap~\eqref{eq:chi_definition} features used below for the isotropic and cubic spectra. Since $\nu_-=k_w$ and $\nu_+=k_w+2k_c$, either scale-free quantity fixes $k_c/k_w$, while the spectral scale fixes the absolute stiffness scale. If all $\{k_j\}$ are independent, however, there is one stiffness scale and two ratios, but two eigenvalues provide only one scale and one scale-free constraint. As in RUS, one ratio degree of freedom remains undetermined unless additional admissible structure or spectral information is supplied.

The same lessons carry over to RUS, but spring constants $k_j$ are effective stiffnesses rather than intrinsic elastic constants. For example, the axial stiffness $k=YA/L$, with $Y$ the Young's modulus, already folds geometry into the discrete parameter. In RUS, this separation is explicit: sample dimensions enter the \RR{} matrices, while the elasticity tensor encodes the intrinsic response. RUS inversion is typically overdetermined, since many resonances estimate only two isotropic or three cubic elastic constants, but it remains nonlinear because mode shapes depend on the material parameters. Thus RUS inversion is a constrained inverse problem with physical structure fixed by elasticity, stability, symmetry, and sample geometry.

\subsection{Forward problem: elastic constants to resonances}
\label{subsec:forward_map}

We write $\Cten$ for the fourth-order elasticity tensor and $C_{ijkl}$ for its components in a fixed orthonormal basis, i.e., $\Cten=(C_{ijkl})_{1\leqslant i,j,k,l \leqslant3}$. Voigt notation represents this symmetric tensor in matrix form by mapping symmetric pairs of Cartesian indices to a single Voigt index, $(11,22,33,23,13,12)\mapsto(1,2,3,4,5,6)$, so that $C_{ijkl}$ are reshaped into a $6\times6$ Voigt matrix $\mat{C}^{\rm V}=(C_{IJ})_{1\leqslant I,J \leqslant6}$~\cite{Nye1985, Landau1986Elasticity, Sokolnikoff1956}. Crystal symmetry enters by imposing additional restrictions on $\Cten$. For isotropic media, such restrictions leave two independent elastic constants, which can be taken as the bulk $K$ and shear $G$ moduli. For cubic solids, they leave three independent constants, $C_{11}$, $C_{12}$, and $C_{44}$. We denote by $\Cphys^{(\mathcal S)}$ the set of thermodynamically stable elasticity tensors compatible with the symmetry class $\mathcal S$. This set is the parameter space for both the forward and inverse problems.

For a Hookean body occupying a volume $V$ with density $\rho(\vect r)$, geometric parameters $L$, elasticity tensor $\Cten$, and crystal symmetry class $\mathcal S$, the forward problem is to compute the resonance frequencies from the data $(\rho,L,\Cten,\mathcal S)$. We use $L$ as a collective label for the sample geometry; for the rectangular parallelepipeds used here, $L=(L_x,L_y,L_z)$.

In the \RR{} method, the displacement field is restricted to a finite-dimensional variational subspace $\Vspace_N\subset H^1(V;\mathbb R^3)$, where $H^1(V;\mathbb R^3)$ is the natural energy space for linear elasticity. The integer $N$ denotes the truncation order, and $d_N=\dim\Vspace_N$ is the number of basis coefficients in a trial displacement $\vect u_N(\vect r,t)=\sum_{\alpha=1}^{d_N}a_\alpha(t)\boldsymbol\Phi_\alpha(\vect r)$, with the concrete basis $\{\boldsymbol\Phi_\alpha(\vect r)\}_\alpha$ specified in App.~\ref{app:supplementary_data_benchmarks}.

For small-amplitude free vibrations, the dynamical equations follow from the elastic Lagrangian. Substituting the \RR{} trial field into the kinetic and elastic energy functionals gives a quadratic Lagrangian for the coefficients $\vect a(t)=(a_\alpha(t))_\alpha$. Stationarity then turns the continuum variational problem into a finite-dimensional generalized eigenvalue problem; see App.~\ref{app:rayleigh_ritz_geometry} for details.

In a chosen basis of $\Vspace_N$, the kinetic and elastic energies define two real symmetric matrices, $\Emat(\rho,L)\in\Sym^{d_N}$ and $\Gmat(\Cten,L)\in\Sym^{d_N}$. Here $\Emat(\rho,L)>0$ is the positive-definite mass/Gram matrix, and $\Gmat(\Cten,L)$ is the stiffness matrix generated by the elasticity tensor and the sample geometry. For a thermodynamically stable elasticity tensor, $\Gmat(\Cten,L)$ is positive semidefinite; under free boundary conditions, its null space contains the rigid-body modes. Stationarity of the finite-dimensional Lagrangian with respect to the coefficients $\vect a$ gives the RUS analog of~\eqref{eq:discrete_generalized_eigenproblem},
\begin{equation}
    \Gmat(\Cten,L)\vect a = \omega^2\Emat(\rho,L)\vect a .
    \label{eq:rus_generalized_eigenproblem_main}
\end{equation}
After removing the rigid-body zero modes, the positive eigenvalues give the squared resonance frequencies.
This variational formulation is classical in RUS and related rectangular-parallelepiped resonance methods~\cite{Demarest1971, Ohno1976, Visscher1991NormalModes, Migliori1993RUS, Migliori1997RUSBook, Leisure1997RUS, Balakirev2019RUSToolbox, Geimer2026}.
For homogeneous density, $\rho(\vect r)=\rho$, the mass matrix factors as $\Emat(\rho,L)=\rho\Emat(L),$ and Eq.~\eqref{eq:rus_generalized_eigenproblem_main} may equivalently be written as $\Gmat(\Cten,L)\vect a=\rho\omega^2\Emat(L)\vect a$. 
This is the common form used for homogeneous RUS samples.

Because $\Emat(\rho,L)$ is positive definite, Eq.~\eqref{eq:rus_generalized_eigenproblem_main} can be written as an ordinary symmetric eigenvalue problem.
Defining the symmetric representative
\begin{equation}
    \Amat_N(\Cten,\rho,L)
    =
    \Emat(\rho,L)^{-1/2}
    \Gmat(\Cten,L)
    \Emat(\rho,L)^{-1/2},
    \label{eq:symmetric_operator_main}
\end{equation}
we obtain $\Amat_N(\Cten,\rho,L)\vect b=\omega^2\vect b$, $\vect b=\Emat(\rho,L)^{1/2}\vect a$.
The spectrum of $\Amat_N(\Cten,\rho,L)$ is the same as that of $\Gmat(\Cten,L)$. We write $\operatorname{spec}_{+,n_\omega}[\Amat_N]$ for the $n_\omega$ positive eigenvalues selected by the experimental window or computational protocol. With this notation, the finite-dimensional forward problem defines a \emph{forward} map. At fixed sample data $(\rho,L,\mathcal S)$, this map is
\begin{equation}
\begin{aligned}
    \mathcal F_N(\,\cdot\,;\rho,L,\mathcal S):
    \Cphys^{(\mathcal S)}
    &\longrightarrow
    \mathbb R_+^{\,n_\omega},
    \\
    \Cten
    &\longmapsto
    \nu = \operatorname{spec}_{+,n_\omega}\!\left[ \Amat_N(\Cten,\rho,L) \right].
\end{aligned}
\label{eq:forward_frequency_map_main}
\end{equation}
In~\eqref{eq:forward_frequency_map_main}, $\nu$ denotes the selected computed squared frequencies, with components $\nu=(\omega_j^2)_{1\leqslant j\leqslant n_\omega}$. The integer $n_\omega$ denotes the number of measured or selected resonances used in the inversion. It is determined by the RUS setup and data-processing choice, such as the measured frequency window and the number of confidently identified peaks; it is independent of the \RR{} basis dimension $d_N$. The entries of $\nu$ are treated as ordered resonance data, not as intrinsically labeled modes. Degeneracies and near-degeneracies may arise from sample geometry, crystal symmetry, or both. Measured RUS data usually consist of resonance frequencies rather than mode shapes or eigenvectors.

For the learning task defined below, it is useful to separate the trivial dependence on absolute sample size from the dependence on shape and elastic constants. For a geometry described by three independent length scales, such as rectangular parallelepipeds with side lengths $(L_x,L_y,L_z)$, one can introduce an overall length scale $R$ and two dimensionless shape coordinates $(\eta,\beta)$. A convenient choice is
\begin{equation}
    R=\sqrt{L_x^2+L_y^2+L_z^2}, \quad m=\int_V\rho(\vect r)\,dV,
\end{equation}
with $m=\rho V$ for homogeneous samples. Here $m$ denotes the total sample mass. For each selected resonance, we define the size-normalized variable
\begin{equation}
    \lambda_j=\frac{m\omega_j^2}{R}. 
    \label{eq:lambda_normalized_main}
\end{equation}
This normalization does not give full scale invariance, but it uses the homogeneity of the forward problem to separate the elastic scale from the dimensionless spectral shape. Using shape coordinates $(\eta,\beta)$, a \emph{reduced forward} map may be written as
\begin{equation}
\begin{aligned}
    \widehat{\mathcal F}_N(\,\cdot\,;\eta,\beta,\mathcal S):
    \Cphys^{(\mathcal S)}
    &\longrightarrow
    \mathbb R_+^{\,n_\omega},
    \\
    \Cten
    &\longmapsto
    \lambda.
\end{aligned}
\label{eq:reduced_forward_map_main}
\end{equation}
In~\eqref{eq:reduced_forward_map_main}, $\lambda$ is the selected size-normalized spectrum defined by~\eqref{eq:lambda_normalized_main}.
This reduction is not restricted to parallelepipeds; it applies whenever the geometry is specified by an overall scale and two independent dimensionless aspect-ratio parameters.
The mathematical properties of Eqs.~\eqref{eq:rus_generalized_eigenproblem_main}--\eqref{eq:reduced_forward_map_main} and the \RR{} method are summarized in App.~\ref{app:rayleigh_ritz_geometry}.

\subsection{Inverse problem: resonances to elastic constants}
\label{subsec:inverse_map}

The inverse problem in RUS is to recover the elasticity tensor from measured resonance frequencies and sample data $(\rho,L,\mathcal S)$. At fixed \RR{} dimension, the measured data give part of the positive spectrum of the symmetric representative $\Amat_N(\Cten,\rho,L)$. The unknown tensor, however, is not arbitrary: it must belong to the physically admissible set determined by crystal symmetry and thermodynamic stability. Thus the inverse problem is finite-dimensional, spectral, and constrained.

The geometric nonuniqueness of the unrestricted problem is immediate.
For a full spectrum $\nu=(\nu_1,
\ldots,\nu_{d_N})$, all real symmetric matrices with that spectrum form the isospectral orbit
\begin{equation}
    \Oiso_\nu
    =
    \left\{
    \Qmat\operatorname{diag}(\nu)\Qmat^T:
    \Qmat\in O(d_N)
    \right\},
    \label{eq:isospectral_orbit_main}
\end{equation}
where $O(d_N)=\left\{\Qmat\in\mathbb R^{d_N\times d_N}:\Qmat^T\Qmat=\Qmat\Qmat^T=\mat{I}\right\}$ is the orthogonal group of $d_N\times d_N$ matrices.
Thus the same spectrum is realized by infinitely many matrices related by orthogonal conjugation.
This is the finite-dimensional analog of the nonuniqueness familiar in inverse eigenvalue problems~\cite{Friedland1979, BoleyGolub1988, Chu1998, Maciazek2022, Kudryavtsev2017, Wu2020}.
However, most matrices in $\Oiso_\nu$ are not physically admissible RUS matrices: they do not arise from a Hookean elasticity tensor through the \RR{} construction, and even if they do, the corresponding elasticity tensor must still satisfy symmetry and stability constraints.

In the idealized case of complete spectral data, the spectrum defines an isospectral orbit in the space of symmetric \RR{} matrices. In RUS, one usually has only a finite set of measured resonances, so the data should be viewed as spectral constraints rather than as a complete orbit. Hookean elasticity, symmetry, and stability define a low-dimensional physical set inside the ambient matrix space; the inverse problem in RUS is to find the physical point, or best approximate physical point, compatible with those constraints.

The stability conditions for $\Cphys^{(\mathcal S)}$ depend on the symmetry class. For cubic materials, the three independent elastic constants satisfy the Born stability conditions
\begin{equation}
    C_{11}+2C_{12}>0,
    \quad
    C_{11}-C_{12}>0,
    \quad
    C_{44}>0.
    \label{eq:cubic_born_conditions_main}
\end{equation}
For isotropic media, stability can be written in terms of the bulk modulus $K$ and shear modulus $G$ as
\begin{equation}
    K>0,
    \quad
    G>0.
    \label{eq:isotropic_born_conditions_main}
\end{equation}
The cubic and isotropic inequalities are the corresponding special cases of the stability criteria summarized by Mouhat and Coudert~\cite{Mouhat2014ElasticStability}.

For fixed sample data, \RR{} basis and dimension, define
\begin{equation}
    \Psi_N:\Cphys^{(\mathcal S)}\to\Sym^{d_N},
    \quad
    \Psi_N(\Cten)=\Amat_N(\Cten,\rho,L).
    \label{eq:psi_map_main}
\end{equation}
Thus $\Psi_N$ is the \RR{} forward representative: it associates a physical elasticity tensor with the symmetric matrix whose eigenvalues are the \RR{} squared resonance frequencies. The forward map~\eqref{eq:forward_frequency_map_main} is obtained by selecting the first $n_\omega$ positive eigenvalues of $\Psi_N$,
\[
    \mathcal F_N(\,\cdot\,;\rho,L,\mathcal S)
    =
    \operatorname{spec}_{+,n_\omega}\circ\Psi_N .
\]
Conversely, the inverse problem uses the same forward representative: admissible tensors are compared by evaluating $\operatorname{spec}_{+,n_\omega}[\Psi_N(\Cten)]$ against the measured data $\hat\nu$.
The image
\begin{equation}
    \Melas_N = \Psi_N\left(\Cphys^{(\mathcal S)}\right) \subset \Sym^{d_N}
    \label{eq:elasticity_manifold_main}
\end{equation}
is the \RR{} elasticity manifold. It is different from the \RR{} variational subspace $\Vspace_N$, which is a space of trial displacement fields; whereas $\Melas_N$ is a set of symmetric matrices generated by varying the physically admissible elasticity tensor.

For fixed symmetry class $\mathcal S$, the map $\Psi_N$ is linear in the components of $\Cten$, but the physical domain is not a vector space due to the fact that stability imposes positivity inequalities. For example, isotropic media require~\eqref{eq:isotropic_born_conditions_main}, while cubic media satisfy~\eqref{eq:cubic_born_conditions_main}. Positive rescalings of a stable tensor remain stable, but arbitrary linear combinations, and in particular sign changes, need not.

\begin{figure*}
    \centering
    \includegraphics*[width=0.4\textwidth]{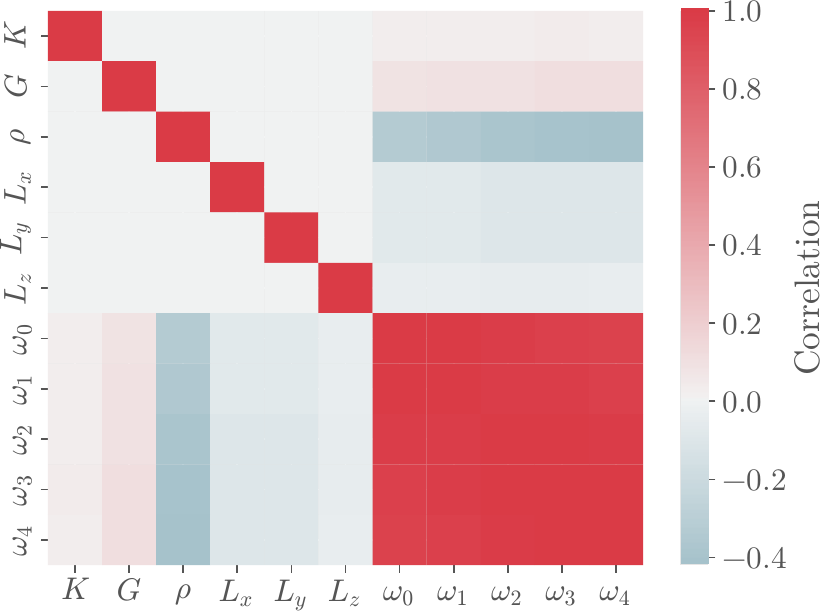}
    \includegraphics*[width=0.4\textwidth]{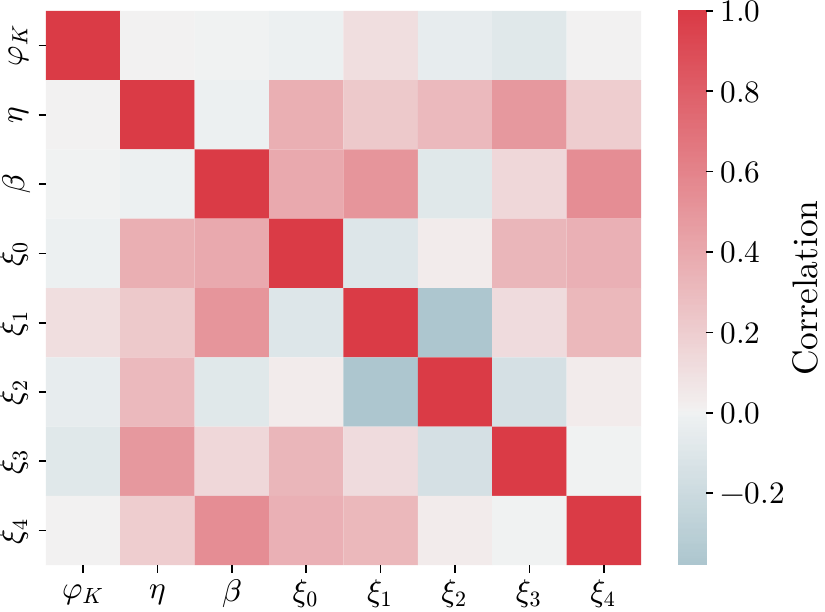}
    \caption{
    Correlation structure before and after the physics-informed reduction. The left panel shows the raw isotropic variables, including sample dimensions, density, and squared resonance-frequency features; the strong correlations among the frequency features reveal substantial redundancy in the raw spectral representation. The right panel shows the reduced isotropic features, built from the aspect-ratio variables $\eta,\beta$ and the scale-free spectral ratios $\xi_j$. The transformed variables exhibit substantially weaker pairwise correlations, indicating that the dominant raw collinearity has been removed while retaining spectral-shape information.
    }
    \label{fig:raw_reduced_correlations}
\end{figure*}

The inverse problem can now be stated in analogy with the retained spectral data produced by the forward problem. Let $\hat\nu=(\hat\nu_1,\ldots,\hat\nu_{n_\omega})$, with $\hat\nu_j=\hat\omega_j^2$, denote the measured squared resonance frequencies used in the inversion. At the same fixed sample data $(\rho,L,\mathcal S)$, the \emph{inverse} map is a constrained isospectral reconstruction map
\begin{equation}
\begin{aligned}
    \mathcal I_N(\,\cdot\,;\rho,L,\mathcal S):
    \mathbb R_+^{\,n_\omega}
    &\longrightarrow
    \Cphys^{(\mathcal S)},
    \\
    \hat\nu
    &\longmapsto
    \Cten_\ast(\hat\nu;\rho,L,\mathcal S).
\end{aligned}
\label{eq:inverse_map_domain_codomain_main}
\end{equation}
Here $\Cten_\ast$ denotes the physical elasticity tensor selected by the inverse procedure. The exact preimage of $\hat\nu$ under the forward map may contain more than one admissible $\Cten$, or may even be empty for noisy data. Thus $\mathcal I_N$, as defined in~\eqref{eq:inverse_map_domain_codomain_main}, is not a strict algebraic inverse of $\mathcal F_N$; it is a physically constrained reconstruction rule that selects one admissible estimate from the data.

In analogy with the reduced forward map~\eqref{eq:reduced_forward_map_main}, at fixed shape $(\eta,\beta)$ and symmetry class $\mathcal S$, the \emph{reduced inverse} map is
\begin{equation}
\begin{aligned}
    \widehat{\mathcal I}_N(\,\cdot\,;\eta,\beta,\mathcal S): \mathbb R_+^{\,n_\omega} 
    &\longrightarrow 
    \Cphys^{(\mathcal S)},
    \\
    \hat\lambda
    &\longmapsto
    \Cten_\ast(\hat\lambda;\eta,\beta,\mathcal S).
\end{aligned}
\label{eq:reduced_inverse_map_main}
\end{equation}
Here $\hat\lambda$ is the measured counterpart of the size-normalized spectrum $\lambda$~\eqref{eq:lambda_normalized_main}. The physics-informed pipeline introduced in Sec.~\ref{sec:physics_informed_reduction} implements this map by separating elastic-ratio prediction from analytic scale recovery.

We obtain $\Cten_\ast$ by minimizing the spectral mismatch (the difference between computed and measured frequencies) over $\Cphys^{(\mathcal S)}$, as detailed in~\eqref{eq:app_constrained_optimization_C}. That is, the selected computed spectrum $\operatorname{spec}_{+,n_\omega}[\Amat_N(\Cten,\rho,L)]$ is compared with the measured data $\hat\nu$, with the search space restricted to $\Cten\in\Cphys^{(\mathcal S)}$. The spectral distance may be a weighted least-squares error when the resonances are ordered and assigned, or a distance between finite multisets when their correspondence is uncertain. In App.~\ref{app:rayleigh_ritz_geometry} we give the equivalent geometric formulation in terms of the elasticity manifold and the isospectral orbit, showing how the constraint $\Cten\in\Cphys^{(\mathcal S)}$ restricts the otherwise nonunique inverse spectral reconstruction. Bayesian, deterministic, and machine-learning approaches to related RUS inverse problems have been developed in several contexts~\cite{RUS2020Isotropic, Bernard2015, Rossin2021, Shi2025InverseRUS, Fukuda2023DLRUS, Yang2022, Liu2023}.

We approximate the reduced inverse map~\eqref{eq:reduced_inverse_map_main} with a feed-forward neural network. The network predicts elastic ratios from the reduced spectrum and sample shape. Section~\ref{subsec:stability_pipeline} defines these ratios and shows how the elastic scale and final constants are recovered.

\section{Physics-informed reduction of the inverse problem}
\label{sec:physics_informed_reduction}

\subsection{Spectral redundancy and reduced coordinates}
\label{subsec:spectral_redundancy}

Section~\ref{sec:rus_inverse_spectral} describes RUS inversion as a constrained inverse-isospectral problem: measured resonances define spectral constraints, while Hookean elasticity, crystal symmetry, and thermodynamic stability restrict the physical operators to the \RR{} elasticity manifold $\Melas_N$. Directly learning the inverse map from raw variables obscures this structure. The raw inputs contain sample dimensions, density or mass information, and resonance frequencies, while the targets are constrained elastic constants. We first identify the variables needed for the regression.

The need for such a reduction is already visible in the raw data. In the initial isotropic dataset, we sampled $(\Cten,\rho,L)$ and computed the resonances with the forward model. The raw inverse inputs $(\rho,L,\{\omega_j\})$ showed strong correlations among the squared frequencies. The sampled physical domain used for the synthetic data is summarized in App.~\ref{app:supplementary_data_benchmarks}. The left panel of Fig.~\ref{fig:raw_reduced_correlations} shows the corresponding correlation matrix. The high correlations among the frequency features indicate that the raw spectrum contains substantial redundant information. This observation motivates the central strategy of our work: rather than asking a model to infer the relevant physical structure from highly correlated raw data, we first transform the variables so that the dominant redundancies are removed.

Several physical effects contribute to this redundancy. At the simplest level, a common elastic scale rescales the spectrum collectively: if the elastic constants are rescaled while their ratios and the sample dimensions are held fixed, the eigenvalues of the \RR{} problem rescale coherently. In addition, changing the sample size modifies the entire boundary-value problem. At fixed geometry, many resonance frequencies depend on only two isotropic or three cubic elastic constants, which also makes them correlated. We return to these mechanisms in Sec.~\ref{subsec:physical_interpretation}; here we use them to motivate the reduced variables.

The first reduction separates absolute size from aspect ratio. We use the size-normalized variables introduced in Sec.~\ref{subsec:forward_map}, $\lambda_j=m\omega_j^2/R$, with $R^2=L_x^2+L_y^2+L_z^2$ for a rectangular parallelepiped. This normalization removes the trivial dependence of the spectrum on a uniform rescaling of the sample dimensions. The remaining geometric dependence is encoded by replacing the three lengths $(L_x,L_y,L_z)$ by an overall scale $R$ and two shape variables $\eta$ and $\beta$. For a rectangular parallelepiped, these variables may be introduced through
\begin{equation}
\begin{aligned}
    L_x &= R\sin(\eta/2)
              \cos(\beta/4),
              \\
    L_y &= R\sin(\eta/2)
              \sin(\beta/4), 
              \label{eq:eta_beta_geometry}
              \\
    L_z &= R\cos(\eta/2).
\end{aligned}
\end{equation}
To remove permutations representing the same aspect ratio, we order the side lengths with $L_z$ largest and $L_y$ smallest. This assigns each aspect ratio to a unique point in the allowed $(\eta,\beta)$ domain, removing the main permutation redundancy among equivalent shapes. The transformation reduces the geometric input from three lengths to two dimensionless shape coordinates, while the absolute scale is handled separately through the size-normalized spectrum~\eqref{eq:lambda_normalized_main}.

\begin{figure}
    \centering
    \includegraphics*[width=0.75\columnwidth]{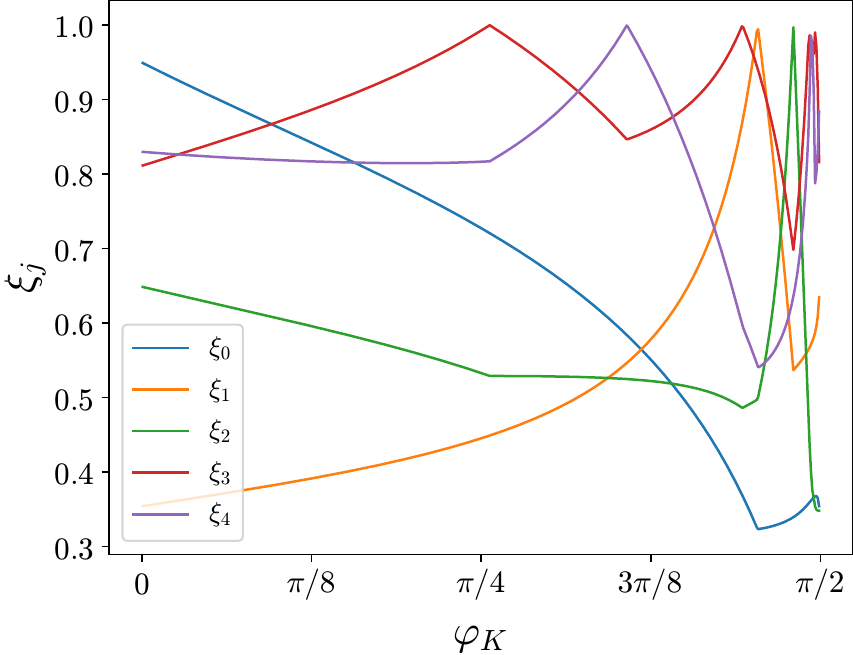}
    \caption{
    Reduced isotropic spectral ratios as functions of $G/K$, expressed through the angular coordinate~\eqref{eq:isotropic_angle_definition}. The ratios $\xi_0,\ldots,\xi_4$ are shown for fixed shape coordinates $(\eta,\beta)=(0.31,0.5)\pi$. The ratios occupy different ranges and respond differently as $G/K$ changes. This illustrates how the scale-free ratios retain complementary spectral-shape information after the common elastic magnitude has been removed.
    }
    \label{fig:isotropic_spectral_ratios} 
\end{figure}

The second reduction removes an extra elastic-scale dependence. With zero-based indexing, the isotropic features are consecutive ratios, the analog of scale-free spring ratios~\eqref{eq:two_mass_scale_free_features},
\begin{equation}
    \xi_j=\frac{\lambda_j}{\lambda_{j+1}}, 
    \qquad
    j=0,\ldots,n_\omega-2. 
    \label{eq:xi_definition}
\end{equation}
For the isotropic benchmarks, $n_\omega=6$, giving the five ratios $\xi_0,\ldots,\xi_4$. At fixed shape, these ratios depend on the modulus ratio $G/K$ but not on the overall elastic magnitude. Figure~\ref{fig:isotropic_spectral_ratios} shows the first five ratios at the representative shape $\eta=0.31\pi$ and $\beta=0.5\pi$. Their weaker correlations across the full dataset are shown in the right panel of Fig.~\ref{fig:raw_reduced_correlations}. The five curves do not collapse onto one another: they occupy different ranges and respond differently as $G/K$ changes. Thus the transformation replaces the strongly scale-correlated raw frequency coordinates by several scale-free spectral-shape features that carry complementary information about the elastic ratio.

For cubic materials, we use normalized spectral spacings analogous to the scale-free gap in~\eqref{eq:two_mass_scale_free_features}. With the convention $\lambda_{-1}\equiv0$,
\begin{equation}
    \chi_j=\frac{\lambda_j-\lambda_{j-1}}{\lambda_{n_\omega-1}},
    \qquad
    j=0,\ldots,n_\omega-1.
    \label{eq:chi_definition}
\end{equation}
For the cubic benchmarks, $n_\omega=20$, giving the twenty spacings $\chi_0,\ldots,\chi_{19}$.
The feature sets $\{\xi_j\}$ and $\{\chi_j\}$ retain local spectral-shape information while removing the overall elastic magnitude. The corresponding correlation matrix for the cubic dataset is shown in App.~\ref{app:supplementary_data_benchmarks}. Neither feature set is statistically independent, since each spectrum comes from a single eigenvalue problem. Both remove the common scale shared by the raw frequencies.

\subsection{Stability-respecting parameterization and inverse pipeline}
\label{subsec:stability_pipeline}

The reduction of the spectral input must be accompanied by an analogous reduction of the target space. Because the elastic constants are constrained by crystal symmetry and thermodynamic stability, direct prediction would require the model to learn both the inverse spectral relation and the admissible stability domain. We instead encode stability before learning by introducing bounded coordinates on the physically admissible target space. In this representation, the neural network predicts only dimensionless angular variables, while the absolute elastic scale is recovered analytically from the homogeneity of the forward problem.






\begin{table}[b]
    \caption{
    Physics-informed transformations used to reduce the inverse problem before the regression model is introduced.
    }
    \label{tab:physics_informed_transformations}
    \begin{ruledtabular}
    \begin{tabular}{lll}
        Raw quantity & Reduced quantity & Physical role \\
        \hline
        $L_x,L_y,L_z$
        & $R,\eta,\beta$
        & Size/shape split \\

        $\omega_j$
        & $\lambda_j=m\omega_j^2/R$
        & Size scaling \\

        $\lambda_j$ (isotropic)
        & $\xi_j=\lambda_j/\lambda_{j+1}$
        & Elastic scaling \\

        $\lambda_j$ (cubic) & $\chi_j=(\lambda_j-\lambda_{j-1})/\lambda_{n_\omega-1}$
        & Spectral gaps \\

        $K,G$
        & $(K,G)\to(M,\phi_K)$
        & Stability coords \\

        $C_{11},C_{12},C_{44}$
        & $(\mathcal K,a,\mu)\to(M,\phi_{\mathcal K},\phi_a)$
        & Stability coords \\
    \end{tabular}
    \end{ruledtabular}
\end{table}

\begin{figure*}
    \centering
    \includegraphics*[width=0.95\textwidth]{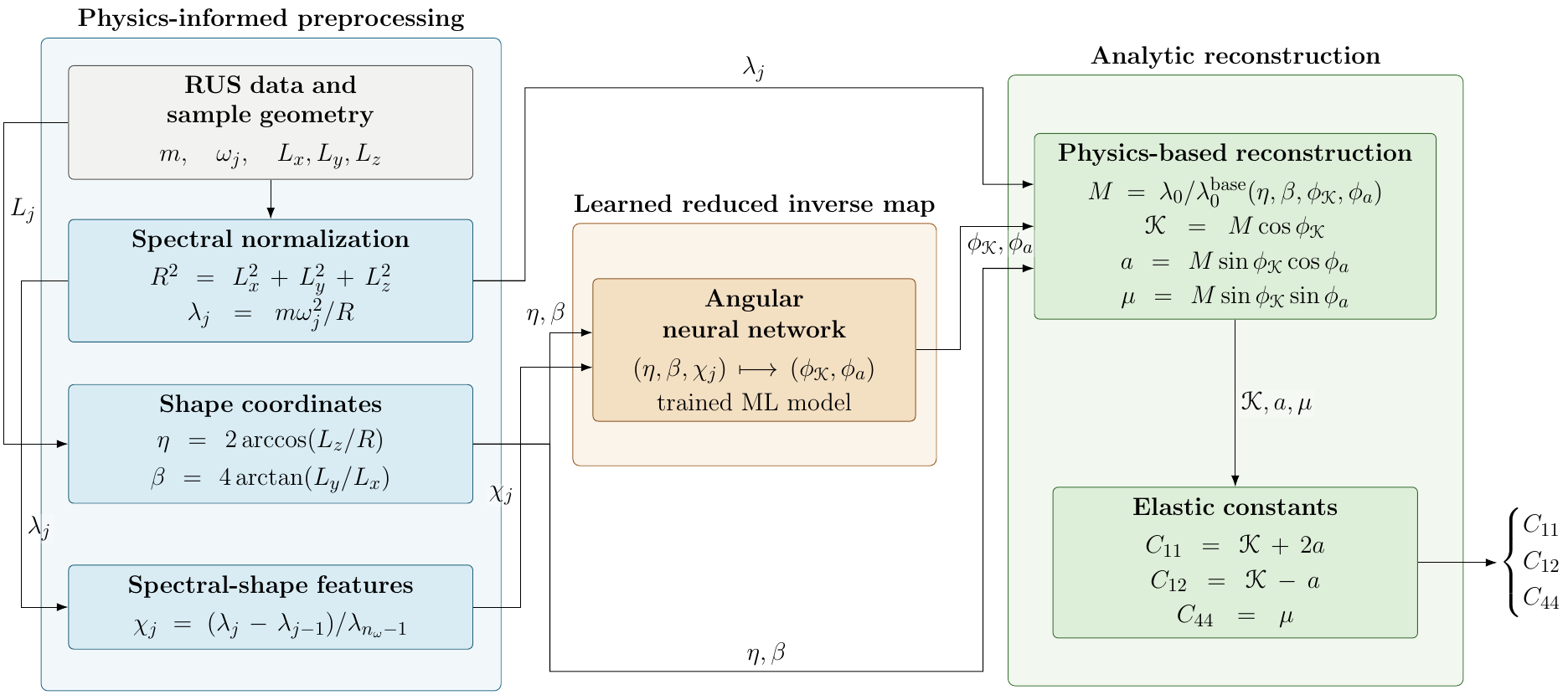}
    \caption{
    Physics-informed inverse-problem pipeline for cubic elastic-constant reconstruction. Raw resonance, mass, and geometry data are transformed into size-normalized spectral variables, scale-free spectral features, and shape coordinates. The feed-forward neural network predicts only the bounded angular variables $(\phi_{\mathcal K},\phi_a)$. A reference forward calculation fixes the elastic scale, after which the final constants $C_{11}$, $C_{12}$, and $C_{44}$ are reconstructed analytically.
    }
    \label{fig:inverse_pipeline}
\end{figure*}

For isotropic materials, the stability domain~\eqref{eq:isotropic_born_conditions_main} is described by the positive bulk modulus $K$ and the shear modulus $G$. We separate their overall magnitude from their ratio by defining
\begin{equation}
    M=\sqrt{K^2+G^2},
    \quad
    \phi_K=\arctan\!\left(\frac{G}{K}\right).
    \label{eq:isotropic_angle_definition}
\end{equation}
The angle $\phi_K\in(0,\pi/2)$ is the reduced target learned by the model. Once $\phi_K$ has been predicted, the moduli are reconstructed from
\begin{equation}
    K=M\cos\phi_K,
    \quad
    G=M\sin\phi_K.
    \label{eq:isotropic_reconstruction}
\end{equation}
The magnitude $M$ is not learned directly. It is determined after prediction by solving a reference forward problem at fixed magnitude and using the proportionality of the normalized eigenvalues with the elastic scale. This is the same scale-separation argument used to define the reduced spectral variables in Sec.~\ref{subsec:spectral_redundancy}.

For cubic solids, the independent elastic constants are $C_{11}$, $C_{12}$, and $C_{44}$, but the stability inequalities~\eqref{eq:cubic_born_conditions_main} are more naturally expressed in terms of the positive variables
\begin{equation}
    \mathcal K=\frac{1}{3}(C_{11}+2C_{12}),
    \quad
    a=\frac{1}{3}(C_{11}-C_{12}),
    \quad
    \mu=C_{44}.
    \label{eq:cubic_stability_variables_sec3}
\end{equation}
In these variables, the cubic Born stability conditions~\eqref{eq:cubic_born_conditions_main} take the simple form
\begin{equation}
    \mathcal K>0,
    \quad
    a>0,
    \quad
    \mu>0.
    \label{eq:cubic_stability_positive_sec3}
\end{equation}
Thus, the allowed cubic target space is mapped to the positive octant in $(\mathcal K,a,\mu)$. This makes stability explicit and connects the learning variables directly to the admissible set $\Cphys^{(\mathcal S)}$ introduced in Sec.~\ref{subsec:inverse_map}.

As in the isotropic case, we separate elastic scale from elastic ratios. Define
\begin{equation}
    M=\sqrt{\mathcal K^2+a^2+\mu^2}.
    \label{eq:cubic_magnitude_sec3}
\end{equation}
The ratios among the three positive variables are represented by two angular coordinates, $\phi_{\mathcal K}$ and $\phi_a$, through
\begin{align}
    \mathcal K &= M\cos\phi_{\mathcal K}, \nonumber\\
    a &= M\sin\phi_{\mathcal K}\cos\phi_a, \label{eq:cubic_angular_reconstruction_sec3}\\
    \mu &= M\sin\phi_{\mathcal K}\sin\phi_a . \nonumber
\end{align}
For positive $\mathcal K$, $a$, and $\mu$, the angles lie in the first octant, $0<\phi_{\mathcal K}<\pi/2$ and $0<\phi_a<\pi/2$. The neural network predicts the bounded targets $(\phi_{\mathcal K},\phi_a)$, while the magnitude $M$ is recovered analytically from the normalized spectrum. The original cubic elastic constants are then reconstructed as
\begin{equation}
    C_{11}=\mathcal K+2a,
    \quad
    C_{12}=\mathcal K-a,
    \quad
    C_{44}=\mu .
    \label{eq:cubic_constants_reconstruction_sec3}
\end{equation}
Equations~\eqref{eq:cubic_stability_variables_sec3}-\eqref{eq:cubic_constants_reconstruction_sec3} show that every reconstructed elasticity tensor satisfies the cubic stability inequalities by construction, provided the predicted angles remain in their physical ranges. They make explicit the low-dimensional structure of the admissible elastic-constant space $\Cphys^{(\mathcal S)}$.

\begin{figure*}[t]
    \centering
    \includegraphics*[height=0.24\textwidth]{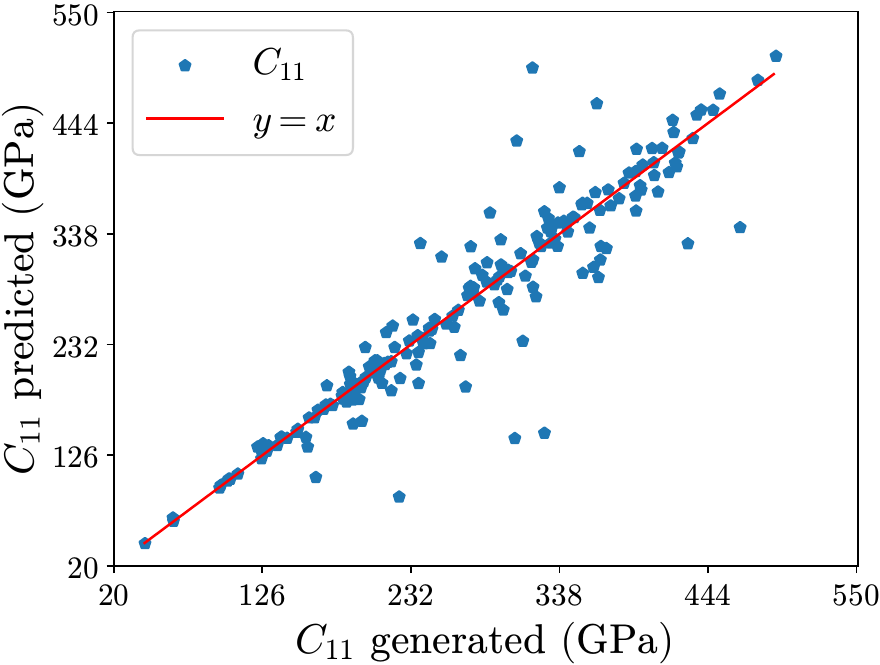}
    \includegraphics*[height=0.24\textwidth]{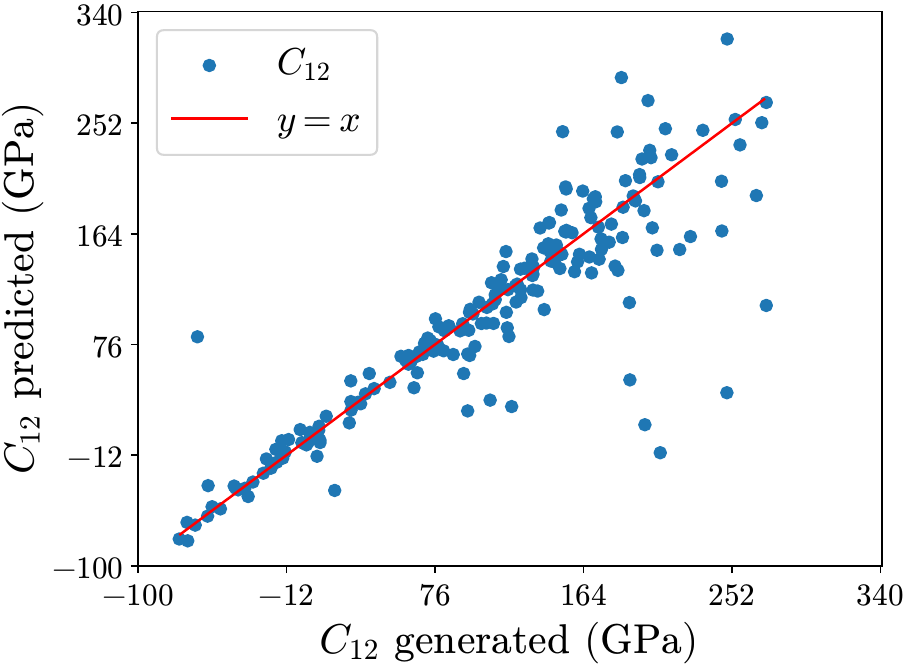}
    \includegraphics*[height=0.24\textwidth]{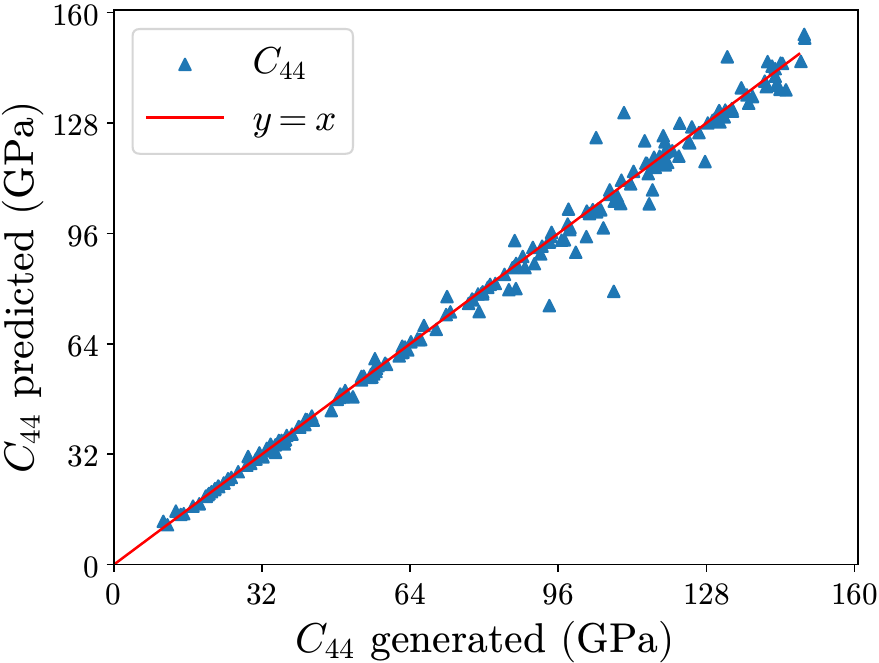}
    \caption{
    Parity plots for the full cubic benchmark in the inverse pipeline. Each panel compares generated values on the horizontal axis with reconstructed values on the vertical axis for one independent cubic elastic constant; the diagonal line represents perfect agreement. In this benchmark, the sample dimensions vary within the domain used by the physics-informed parameterization, while the elastic constants span the thermodynamically stable cubic domain. The agreement is strongest for $C_{44}$ and broadest for $C_{12}$. The broader spread of the $C_{12}$ distribution reflects the sensitivity of $C_{12}=\mathcal K-a$ to propagated errors in the reconstructed variables $\mathcal K$ and $a$.
    }
    \label{fig:reported_predicted_constants}
\end{figure*}

Table~\ref{tab:physics_informed_transformations} summarizes the transformations used in the reduced inverse problem, and Fig.~\ref{fig:inverse_pipeline} shows how they enter the cubic inverse pipeline, which consists of three stages. Physics-informed preprocessing converts the sample data and measured resonances into shape coordinates and scale-free spectral features. The feed-forward neural network learns the reduced inverse map~\eqref{eq:reduced_inverse_map_main}, predicting the angular elastic-ratio coordinates. Finally, a reference forward calculation recovers the magnitude $M$, and the elastic constants are reconstructed analytically from~\eqref{eq:cubic_constants_reconstruction_sec3}. The results are reported in Sec.~\ref{sec:results_discussion}.

\section{Results and discussion}
\label{sec:results_discussion}

\subsection{Isotropic and cubic benchmarks}
\label{subsec:benchmarks}

After applying the transformations of Sec.~\ref{sec:physics_informed_reduction}, we train neural networks to approximate the reduced inverse map~\eqref{eq:reduced_inverse_map_main}. The isotropic model predicts $\phi_K$ from $(\eta,\beta,\{\xi_j\})$, while the cubic model predicts $(\phi_{\mathcal K},\phi_a)$ from $(\eta,\beta,\{\chi_j\})$. Figure~\ref{fig:inverse_pipeline} summarizes the cubic pipeline.

We report two complementary results. The full benchmark tests the scope of the physics-informed parameterization by allowing the sample dimensions, and therefore the aspect ratios, to vary within the generated domain while sampling the thermodynamically stable elastic region. A fixed-geometry benchmark, using the $3{:}4{:}5$ aspect ratio~\cite{Fukuda2023DLRUS}, is used as a controlled comparison in which the aspect-ratio dependence of the inverse map is removed. This separation is useful because the full benchmark probes the generality of the reduced coordinates, whereas the fixed-geometry benchmark isolates the accuracy of the elastic-ratio reconstruction in a simpler setting.
Datasets, network architectures, and angular-model metrics are summarized in App.~\ref{app:supplementary_data_benchmarks}.

As an intermediate check, the reduced angular maps give test MAE values of approximately $0.024$ for $\phi_K$ and $0.028$ for $(\phi_{\mathcal K},\phi_a)$, with corresponding RMSE values of $0.040$ and $0.062$. These metrics show that the reduced maps are learnable; however, the experimentally relevant benchmark is the final reconstruction of the elasticity tensor $\Cten$.

After analytic scale recovery and reconstruction, the full cubic benchmark gives MAE values of $20.37(35.15)~\mathrm{GPa}$ for $C_{11}$, $24.30(41.33)~\mathrm{GPa}$ for $C_{12}$, and $2.13(3.66)~\mathrm{GPa}$ for $C_{44}$. The corresponding parity plots are shown in Fig.~\ref{fig:reported_predicted_constants}. The predictions remain organized around the diagonal, but the dispersion is broader than in the fixed-geometry benchmark (not shown). This is expected: the full benchmark tests a single reduced inverse map across varying sample dimensions, including aspect ratios, and across the thermodynamically stable cubic domain, so aspect-ratio-dependent spectral sensitivity is part of the learning problem, not a fixed parameter.

\begin{figure}
    \centering
    \includegraphics*[height=0.38\columnwidth]{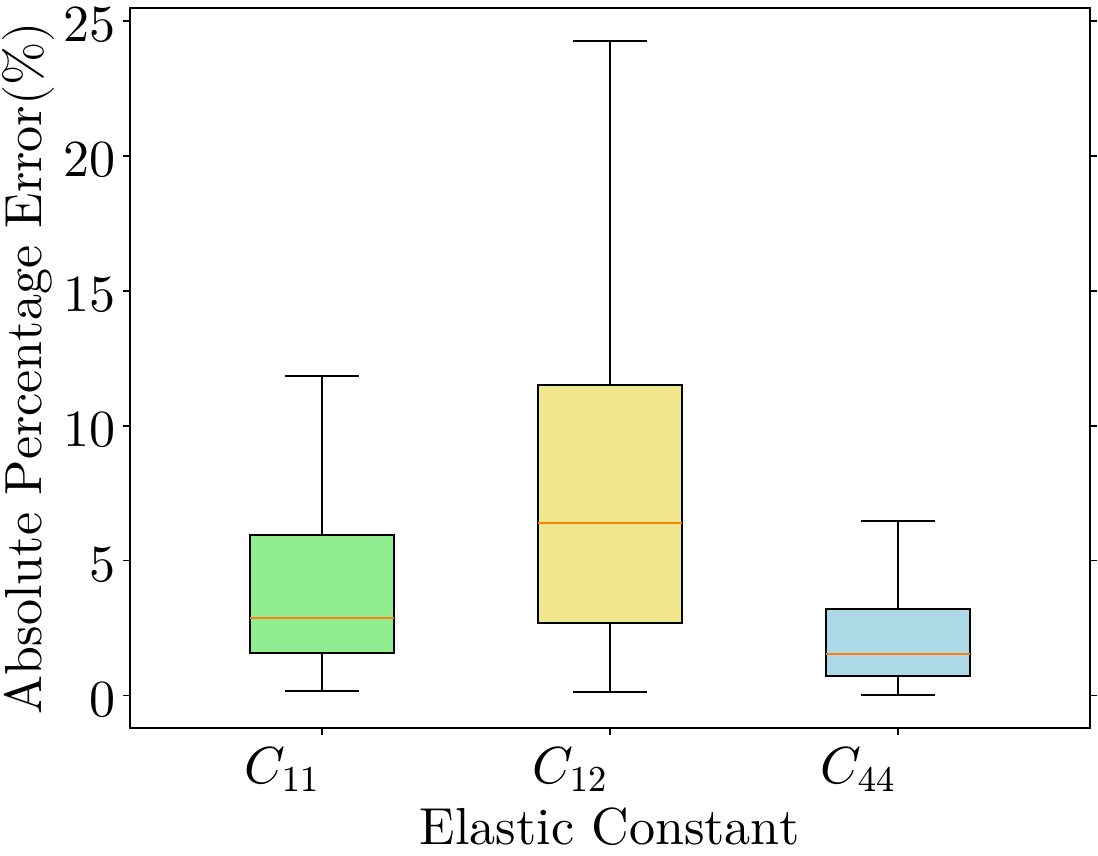}
    \includegraphics*[trim={26pt 0 0 0}, height=0.38\columnwidth]{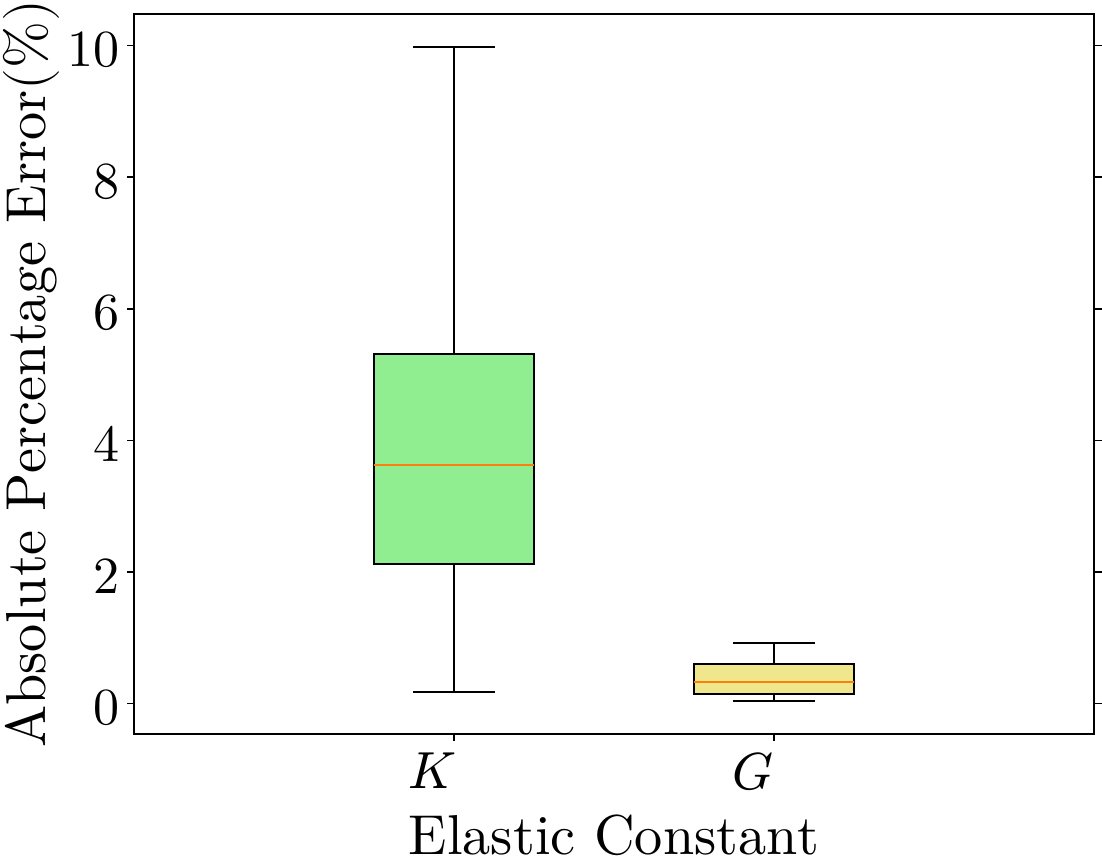}
    \caption{
    Absolute percentage-error distributions for the fixed-geometry benchmarks. The left panel shows the cubic reconstruction errors for $C_{11}$, $C_{12}$, and $C_{44}$; the right panel shows the isotropic reconstruction errors for $K$ and $G$. Both panels use the $3{:}4{:}5$ aspect ratio as a controlled comparison, so that the aspect-ratio dependence present in the full benchmark is removed. The corresponding MAPE values are $4.14(3.87)\%$, $8.31(8.50)\%$, and $2.44(2.86)\%$ for $C_{11}$, $C_{12}$, and $C_{44}$, respectively, and $4.0(3.6)\%$ and $0.4(0.3)\%$ for $K$ and $G$, respectively.
    }
    \label{fig:absolute_error_distribution}
\end{figure}

The fixed-geometry benchmark gives smaller percentage errors because it removes the aspect-ratio dependence of the inverse map. The two benchmarks are complementary: the full benchmark tests the scope of the parameterization, while Fig.~\ref{fig:absolute_error_distribution} shows the percentage-error distributions at fixed geometry. Our approach is parsimonious: it reaches few-percent errors using scalar spectral features and direct regression, without image encoding or separate classification and regression stages. This also reduces the required training data and computational cost. Full-benchmark error distributions are provided in App.~\ref{app:supplementary_data_benchmarks}. The origin of the broader full-benchmark errors is discussed next.

\subsection{Physical interpretation and limitations}
\label{subsec:physical_interpretation}

The benchmark behavior reflects the physical correlations present in the raw inputs.

One redundancy is the common elastic scale. At fixed sample dimensions and fixed elastic ratios, multiplying all elastic constants by a common factor rescales the \RR{} stiffness matrix and hence the eigenvalues coherently. The size-normalized spectrum and the feature sets $\{\xi_j\}$ and $\{\chi_j\}$ remove this common scale from the input while retaining local spectral-shape information. The absolute scale is recovered in the third step of the inverse pipeline of Fig.~\ref{fig:inverse_pipeline}, after the angular variables have been predicted.

A second source is the low-dimensional ratio structure of the stiffnesses. Isotropic and cubic solids have only two and three independent elastic constants, respectively, but the raw spectra hide the relevant ratios behind size, density, aspect-ratio, and elastic-scale effects. The angular targets $\phi_K$ and $(\phi_{\mathcal K},\phi_a)$ expose these ratios directly. In this representation, the network learns bounded elastic-ratio coordinates and not isolated elastic scales.

Overlapping spectral information is also expected physically. Many low-frequency resonances can contain substantial shear deformation, so several modes may probe similar shear-dominated combinations of elastic constants. This is consistent with the benchmark behavior: $C_{44}=\mu$ is reconstructed most accurately, whereas $C_{11}=\mathcal K+2a$ and especially $C_{12}=\mathcal K-a$ combine reconstructed variables and inherit larger propagated errors.

Sample geometry also correlates the spectrum. Changing side-length ratios shifts many modes together and may produce near-degeneracies. In some regions, resonances may also become less sensitive to particular elastic combinations, increasing the spread of the reconstructed constants. The map $(L_x,L_y,L_z)\mapsto(R,\eta,\beta)$ separates overall size from aspect ratio, but the full benchmark must still account for this shape dependence. The resulting percentage-error distributions are shown in Fig.~\ref{fig:app_full_benchmark_ape} of App.~\ref{app:supplementary_data_benchmarks}.

Percentage-error metrics require additional care for $C_{12}$. Since $C_{12}=\mathcal K-a$ can be small or negative in the thermodynamically stable cubic domain, modest absolute errors in $\mathcal K$ or $a$ can translate into large relative errors. For this reason, the full benchmark is summarized primarily with absolute errors, while percentage-error distributions are used as diagnostics. The broad $C_{12}$ distribution is naturally connected to this cancellation-sensitive effect.

The geometric formulation of Sec.~\ref{subsec:inverse_map} summarizes these observations: for fixed sample data and crystal symmetry, the admissible elasticity tensors trace out the \RR{} elasticity manifold $\Melas_N$~\eqref{eq:elasticity_manifold_main} inside the larger space of symmetric \RR{} matrix representatives. The measured RUS spectrum is therefore a spectral representation of a low-dimensional, stability-constrained physical set. The reduced variables provide coordinates for this set.

\section{Conclusions and outlook}
\label{sec:conclusions_outlook}

We have described RUS inversion as a constrained inverse-isospectral problem on the manifold of physically admissible elasticity tensors. In the \RR{} representation, the elasticity tensor, sample geometry, and density determine the resonance frequencies. Crystal symmetry and thermodynamic stability restrict the allowed elastic constants, so the inverse search is confined to a small physical region of the full matrix space.

We use this structure before training. The sample dimensions are separated into an overall size and two shape coordinates, and frequency ratios or normalized gaps remove the elastic scale from the spectral inputs. Positive angular coordinates enforce stability in the isotropic and cubic targets. The neural network learns the reduced inverse map from spectral shape and sample shape to elastic ratios. A reference forward calculation then recovers the elastic magnitude, and the final constants follow from the analytic reconstruction formulas.

The benchmarks show consistent elastic-constant reconstruction with feed-forward regression neural network models. For the full cubic benchmark, the reconstructed constants give MAE values of $20.37(35.15)$, $24.30(41.33)$, and $2.13(3.66)~\mathrm{GPa}$ for $C_{11}$, $C_{12}$, and $C_{44}$, respectively. The fixed-geometry benchmark gives smaller percentage errors, with cubic MAPE values of $4.14(3.87)\%$, $8.31(8.50)\%$, and $2.44(2.86)\%$, and isotropic values of $4.0(3.6)\%$ and $0.4(0.3)\%$. The larger spread for $C_{12}$ follows naturally from the cancellation-sensitive reconstruction $C_{12}=\mathcal K-a$.

The resulting pipeline is parsimonious in both its physical representation and its computational use. It uses scalar shape and spectral features, a direct feed-forward regression, and analytic scale recovery. It avoids spectral images and separate classification and regression stages. The reduced learned task requires less training data and computation than previous approaches while retaining few-percent errors in the fixed-geometry benchmarks.

The present study is limited to isotropic and cubic elasticity and to synthetic spectra generated with the same \RR{} model used for training. These symmetry classes have low-dimensional stability domains. Lower-symmetry crystals have more independent constants and more complicated stability conditions, and will require new stability-respecting coordinates. Experimental applications must also handle uncertain mode assignments, peak overlap, damping, dimensional errors, and differences between the model and the measured sample. A mode-by-mode decomposition into volumetric and shear strain energies could clarify which elastic combinations are most strongly probed by the low-frequency spectrum.

\begin{acknowledgments}
The authors thank P. Giraldo-Gallo, B. Maiorov, and M. Oster for fruitful discussions. 
J.R.\ acknowledges support from the ICTP through the Associates Programme (2026--2031) and from the Office of the Vice President for Research and Creative Activities and the Office of the Vice President for Research of the Faculty of Science at Universidad de los Andes under the FAPA grant.
\end{acknowledgments}

\paragraph*{Code and data availability.} 
The implementation associated with this work is available in the public repository \href{https://github.com/cubos-d/RUSpectroscopy_Tools}{\texttt{RUSpectroscopy\_Tools}}~\cite{Cubillos2026}. The repository contains the software resources needed to reproduce the forward RUS calculations, data generation, reduced inverse pipeline, and benchmark analyses. 

\appendix

\section{\RR{} formulation, stability, and isospectral geometry}
\label{app:rayleigh_ritz_geometry}

This appendix provides the \RR{} framework used in Sec.~\ref{sec:rus_inverse_spectral}. The construction follows standard variational formulations of free elastic vibrations and RUS~\cite{Demarest1971, Ohno1976, Holland1968, Visscher1991NormalModes, Migliori1993RUS, Migliori1997RUSBook, Leisure1997RUS, Balakirev2019RUSToolbox, Geimer2026}. General background on elasticity, energy methods, and vibrations can be found in Refs.~\cite{Landau1986Elasticity, Sokolnikoff1956, Langhaar2016, Thomsen2021, Leisure2017Ultrasonic}. Throughout this appendix, the dependence of $\Emat$, $\Gmat$, and $\Amat_N$ on fixed quantities such as density, geometry, and basis choice is suppressed when no ambiguity arises.

The starting point is the Lagrangian of a freely vibrating elastic body, $L=T-U$, where $T$ is the kinetic energy and $U$ is the elastic energy. The \RR{} method restricts the displacement field to the finite-dimensional trial space $\Vspace_N$. For a bounded elastic body, this replaces the continuum free-vibration spectrum by $d_N=\dim\Vspace_N$ eigenvalues, including the rigid-body zero modes under free boundary conditions. Substitution into the elastic Lagrangian gives a quadratic form in the expansion coefficients, whose stationarity yields the generalized eigenvalue problem~\eqref{eq:rus_generalized_eigenproblem_main}. The inverse pipeline retains only $n_\omega$ positive eigenvalues, as determined by the measured frequency window, peak identification, and data-processing choices.

\paragraph*{\RR{} subspace and mass inner product.}

Let $V\subset\mathbb R^3$ be the reference configuration of an elastic body and let $\Vspace=H^1(V;\mathbb R^3)=W^{1,2}(V;\mathbb R^3)$ be the displacement field space. Here, $W^{k,p}$ is the Sobolev space of fields whose weak derivatives through order $k$ belong to $L^p$, and $H^k=W^{k,2}$. The choice $k=1$ is dictated by linear elasticity, because the strain tensor contains first derivatives of the displacement field. The choice $p=2$ is dictated by the energy: the displacement must be square integrable to give finite kinetic energy, and its first weak derivatives must be square integrable to give finite elastic energy. Thus $H^1=W^{1,2}$ is the natural energy space for the variational formulation.
The \RR{} method replaces $\Vspace$ by a finite-dimensional subspace
\begin{equation}
    \Vspace_N
    =
    \operatorname{span}\{\boldsymbol\Phi_1,\ldots,\boldsymbol\Phi_{d_N}\}
    \subset \Vspace .
    \label{eq:app_rr_space}
\end{equation}
The concrete polynomial basis used to generate the synthetic spectra is specified in App.~\ref{app:supplementary_data_benchmarks}, where the abstract index $\alpha$ is realized as a multi-index labeling Cartesian components and polynomial powers. With this abstract notation, a trial displacement is written as
\begin{equation}
    \vect u_N(\vect r,t) = \sum_{\alpha=1}^{d_N} a_\alpha(t)\boldsymbol\Phi_\alpha(\vect r),
    \label{eq:app_trial_displacement}
\end{equation}
where $\vect a(t)=(a_1(t),\ldots,a_{d_N}(t))^T.$ The mass inner product
\begin{equation}
    \langle \vect u,\vect v\rangle_\rho
    =
    \int_V \rho(\vect r)\,\vect u\cdot\vect v\,dV
    \label{eq:app_mass_inner_product}
\end{equation}
defines the geometry of $\Vspace_N$. In the basis~\eqref{eq:app_rr_space}, this inner product is represented by
\begin{equation}
    (\Emat)_{\alpha\beta}(\rho)
    =
    \int_V
    \rho(\vect r)
    \boldsymbol\Phi_\alpha(\vect r)\cdot
    \boldsymbol\Phi_\beta(\vect r)
    \,dV .
    \label{eq:app_E_matrix}
\end{equation}
The weighted Gram matrix $\Emat$ is real symmetric and positive definite if the basis functions are linearly independent with respect to the mass inner product.
For homogeneous density, $\Emat(\rho,L)=\rho\Emat(L)$ reduces to a standard Gram matrix.

\paragraph*{Elastic bilinear form and stability.}

The linearized strain tensor is $\varepsilon_{ij}(\vect u)=(\partial u_i/\partial r_j+\partial u_j/\partial r_i)/2$. The elasticity tensor $\Cten=(C_{ijkl})$ defines the elastic energy bilinear form
\begin{equation}
    B_{\Cten}(\vect u,\vect v) = \int_V C_{ijkl}\, \varepsilon_{ij}(\vect u)\, \varepsilon_{kl}(\vect v) \,dV . 
    \label{eq:app_elastic_bilinear_form}
\end{equation}
Equivalently, using the minor symmetries of the elasticity tensor, the same bilinear form may be written in terms of displacement gradients.
Its \RR{} matrix representation is
\begin{equation}
    (\Gmat)_{\alpha\beta}(\Cten)
    =
    B_{\Cten}(\boldsymbol\Phi_\alpha,\boldsymbol\Phi_\beta).
    \label{eq:app_Gamma_matrix}
\end{equation}
The matrix $\Gmat(\Cten)$ is real symmetric by the symmetry of the elastic energy.
Mechanical stability requires the elastic energy density to be positive for all nonzero strains compatible with the material symmetry.
For a stable elasticity tensor, $B_{\Cten}(\vect u,\vect u)\ge 0$ on $\Vspace_N$, and therefore $\Gmat(\Cten)$ is positive semidefinite.
The precise inequalities defining the stability domain depend on the crystal class $\mathcal S$; the necessary and sufficient stability conditions for the different crystal systems are summarized in Ref.~\cite{Mouhat2014ElasticStability}.
For cubic materials this gives Eq.~\eqref{eq:cubic_born_conditions_main}.
Stress-strain relations for cubic crystals, tensor notation, and Voigt representations are discussed in Refs.~\cite{Thomas1956CubicStressStrain, Nye1985, Landau1986Elasticity, Sokolnikoff1956}.

\paragraph*{Generalized eigenvalue problem and self-adjointness.}

The continuum kinetic and elastic energies are
\begin{equation*}
    T[\dot{\vect u}] = \frac{1}{2} \int_V \rho(\vect r)\, |\dot{\vect u}(\vect r,t)|^2\,dV,
    \quad
    U[\vect u] = \frac{1}{2} B_{\Cten}(\vect u,\vect u).
\end{equation*}
After substituting the trial field into the Lagrangian $L[\vect u,\dot{\vect u}]=T[\dot{\vect u}]-U[\vect u]$, the finite-dimensional Lagrangian takes the quadratic form
\begin{equation}
    L_N(\vect a,\dot{\vect a}) = \frac{1}{2} \dot{\vect a}^T\Emat(\rho)\dot{\vect a} - \frac{1}{2} \vect a^T\Gmat(\Cten)\vect a .
    \label{eq:app_finite_lagrangian}
\end{equation}
The Euler-Lagrange equations are $\Emat(\rho)\ddot{\vect a} + \Gmat(\Cten)\vect a = 0$.
For harmonic motion $\vect a(t)=\vect a e^{i\omega t}$, this yields the generalized eigenvalue problem~\eqref{eq:rus_generalized_eigenproblem_main}: $\Gmat(\Cten)\vect a = \omega^2\Emat(\rho)\vect a$.
Under free boundary conditions, rigid-body translations and rotations have zero strain energy, producing zero eigenvalues. The physical RUS frequencies are obtained from the positive eigenvalues of~\eqref{eq:rus_generalized_eigenproblem_main}.

Define $T_{\Cten}:\Vspace_N\to\Vspace_N$ by $\langle \vect u,T_{\Cten}\vect v\rangle_\rho=B_{\Cten}(\vect u,\vect v)$. In coordinates, $T_{\Cten}=\Emat^{-1}\Gmat(\Cten)$.
This matrix is not generally symmetric in the Euclidean inner product, but the operator is self-adjoint with respect to the mass inner product:
\begin{equation}
    \langle \vect u,T_{\Cten}\vect v\rangle_\rho
    =
    B_{\Cten}(\vect u,\vect v)
    =
    B_{\Cten}(\vect v,\vect u)
    =
    \langle T_{\Cten}\vect u,\vect v\rangle_\rho .
    \label{eq:app_E_self_adjointness}
\end{equation}
Consequently, the generalized eigenvalues are real and eigenvectors associated with distinct eigenvalues are orthogonal in the mass inner product.

\paragraph*{Symmetric representative and orthogonal diagonalization.}

Since $\Emat$ is symmetric positive definite, it has a unique positive square root $\Emat^{1/2}$. The generalized eigenvalue problem is self-adjoint with respect to the mass inner product, but not generally symmetric in the ordinary Euclidean inner product on coefficient space. To express the same spectrum in Euclidean symmetric form, define
\begin{equation}
    \Amat_N(\Cten)=\Emat^{-1/2}\Gmat(\Cten)\Emat^{-1/2}.
    \label{eq:app_A_operator}
\end{equation}
With $\vect b=\Emat^{1/2}\vect a$, Eq.~\eqref{eq:rus_generalized_eigenproblem_main} becomes
\begin{equation}
    \Amat_N(\Cten)\vect b = \omega^2\vect b .
    \label{eq:app_symmetric_eigenproblem}
\end{equation}
The transformation from $(\Gmat,\Emat)$ to $\Amat_N$ changes the matrix representative but not the generalized eigenvalues. It places the problem in the standard space of real symmetric matrices, where ordinary orthogonal diagonalization applies:
\begin{equation}
    \Qmat^T \Amat_N(\Cten) \Qmat = \operatorname{diag}(\nu_1,\ldots,\nu_{d_N}),
    \quad
    \Qmat\in O(d_N),
    \label{eq:app_orthogonal_diagonalization}
\end{equation}
with $\nu_j=\omega_j^2$. This Euclidean symmetric representative is the one used to formulate the inverse problem in RUS as a constrained isospectral problem.

\paragraph*{Isospectral orbits.}

Let
\begin{equation}
    \Sym^{d_N} = \left\{ \Amat\in\mathbb R^{d_N\times d_N} : \Amat^T=\Amat \right\}
\end{equation}
denote the vector space of real symmetric $d_N\times d_N$ matrices.
Let $\nu=(\nu_1,\ldots,\nu_{d_N})$ be a complete spectrum. The orthogonal group $O(d_N)$ acts on $\Sym^{d_N}$ by conjugation, $\Amat\mapsto\Qmat\Amat\Qmat^T$. The orbit through $\operatorname{diag}(\nu)$ is
\begin{equation}
    \Oiso_\nu
    =
    \left\{
    \Qmat\operatorname{diag}(\nu)\Qmat^T:
    \Qmat\in O(d_N)
    \right\}.
    \label{eq:app_isospectral_orbit}
\end{equation}
This orbit is the set of all real symmetric matrices with spectrum $\nu$. It represents the unrestricted nonuniqueness of reconstruction from spectral data: changing $\Qmat$ changes the eigenvectors while preserving the eigenvalues. 
Thus a spectrum alone does not identify a unique symmetric matrix. This point is central in finite inverse eigenvalue problems and matrix reconstruction from spectral data~\cite{Friedland1979, BoleyGolub1988, Chu1998, Maciazek2022, Kudryavtsev2017, Wu2020}. The physical inverse problem in RUS is much more constrained, because the matrix must be generated by a stable Hookean elasticity tensor through the \RR{} construction.

In the original generalized coordinates, the analogous orbit of stiffness matrices is
\begin{equation}
    \Oiso^\Emat_\nu
    =
    \left\{
    \Emat^{1/2}\Qmat\operatorname{diag}(\nu)\Qmat^T\Emat^{1/2}:
    \Qmat\in O(d_N)
    \right\}.
    \label{eq:app_E_orbit}
\end{equation}
This is the same isospectral freedom expressed in the coordinates associated with the \RR{} mass inner product. The main text uses the Euclidean representative $\Amat_N$ because it makes the isospectral geometry explicit in the standard space $\Sym^{d_N}$.

\paragraph*{Elasticity manifold and constrained inverse problem.}

For fixed density, geometry, basis, and symmetry class, define
\begin{equation}
    \Psi_N:\Cphys^{(\mathcal S)}\to\Sym^{d_N},
    \quad
    \Psi_N(\Cten)=\Emat^{-1/2}\Gmat(\Cten)\Emat^{-1/2}.
    \label{eq:app_Psi_map}
\end{equation}
The image
\begin{equation}
    \Melas_N=\Psi_N\left(\Cphys^{(\mathcal S)}\right)
    \label{eq:app_elasticity_manifold}
\end{equation}
is the \RR{} elasticity manifold in the symmetric-representative picture. It is the set of symmetric matrix representatives, equivalent to the generalized \RR{} eigenvalue problems, generated by thermodynamically stable elasticity tensors in the assumed symmetry class $\mathcal S$. The word ``manifold'' refers to this parameterized physical set in matrix space, not to the \RR{} variational subspace $\Vspace_N$.

\begin{table*}[t]
    \caption{
    Sampling domain used to generate the synthetic RUS datasets. The table summarizes the dimensional and material ranges, angular targets, and retained reduced spectral features used in the isotropic and cubic benchmarks.
    }
    \label{tab:app_sampling_domain}
    \begin{ruledtabular}
    \begin{tabular}{llll}
        Quantity & Symbol & Range or value & Comment \\
        \hline
        Sample lengths
        & $L_x,L_y,L_z$
        & $0.1$--$1~\mathrm{cm}$
        & Variables before $R,\eta,\beta$ conversion \\

        Density
        & $\rho$
        & $0.2$--$10~\mathrm{g}/\mathrm{cm}^{3}$
        & Sample mass computation \\

        Isotropic bulk, shear moduli
        & $K,G$
        & $0.3$--$5.6~\mathrm{Tdyn}/\mathrm{cm}^{2}$
        & Raw data; sampled with $K,G>0$ \\

        Cubic elastic constants
        & $C_{11},C_{12},C_{44}$
        & not sampled directly
        & Born-stable via $\mathcal K,a,\mu>0$ \\

        Angular targets
        & $\phi_K,\phi_{\mathcal K},\phi_a$
        & $0$--$\pi/2$
        & Stability-respecting ratio coordinates \\

        Isotropic spectral features
        & $\xi_0,\ldots,\xi_4$
        & $5$ ratios
        & Consecutive normalized-eigenvalue ratios \\

        Cubic spectral features
        & $\chi_0,\ldots,\chi_{19}$
        & $20$ spacings
        & Retained normalized spectral spacings \\
    \end{tabular}
    \end{ruledtabular}
\end{table*}

Before stability is imposed, the stiffness matrix depends linearly on the elastic constants. A general elasticity tensor has $21$ independent components, reduced by crystal symmetry to two for isotropic elasticity and three for cubic elasticity. Stability selects an open positive cone within this symmetry-restricted parameter space: positive rescalings and positive linear combinations remain admissible, whereas arbitrary linear combinations do not. Stability therefore restricts the domain without changing its dimension. The image $\Melas_N$ under $\Psi_N$ has dimension at most two or three, respectively, inside the ambient matrix space $\Sym^{d_N}$, whose dimension is $d_N(d_N+1)/2$.

In the idealized full-spectrum setting, let $\hat\nu=(\hat\nu_1,\ldots,\hat\nu_{d_N})$ denote a complete spectrum for the symmetric representative. The exact finite-dimensional inverse problem can then be written as the constrained isospectral intersection problem
\begin{equation}
    \text{find } \Amat\in\Melas_N\cap\Oiso_{\hat\nu}.
    \label{eq:app_intersection_problem}
\end{equation}
Equivalently, one may seek the closest point on the elasticity manifold to the corresponding isospectral orbit, 
\begin{equation}
    \min_{\Amat\in\Melas_N} \operatorname{dist}\left(\Amat,\Oiso_{\hat\nu}\right)^2.
    \label{eq:app_geometric_optimization}
\end{equation}
For example, with the Frobenius norm,
$
    \operatorname{dist}\left(\Amat,\Oiso_{\hat\nu}\right) = \inf_{\mat{B}\in\Oiso_{\hat\nu}} \|\Amat-\mat{B}\|_F . 
$
Here the orbit formulation refers to the idealized case in which the full finite-dimensional spectrum is prescribed; when only the experimentally retained resonances are available, the same idea is implemented through the spectral mismatch below.
In practical RUS, the measured data usually consist of only finitely many retained positive resonances, possibly with noise and uncertain mode correspondence. In that case the inverse problem is more naturally written as a constrained spectral mismatch,
\begin{equation}
    \min_{\Cten\in\Cphys^{(\mathcal S)}} d_{\rm spec} \left( \operatorname{spec}_{+,n_\omega}[\Psi_N(\Cten)], \hat\nu \right)^2 + \alpha\mathcal R(\Cten).
    \label{eq:app_constrained_optimization_C}
\end{equation}
Here $\mathcal R(\Cten)$ is an optional regularization term, weighted by $\alpha\geq0$, that can incorporate prior information about the elasticity tensor; setting $\alpha=0$ gives the unregularized spectral fit. The constraints are encoded in the feasible sets. In~\eqref{eq:app_geometric_optimization}, the search is over $\Melas_N$, not over all of $\Sym^{d_N}$. In~\eqref{eq:app_constrained_optimization_C}, the search is over $\Cphys^{(\mathcal S)}$, not over arbitrary elastic constants. Thus~\eqref{eq:app_constrained_optimization_C} is the abstract mathematical form of the inverse problem in RUS used in the main text. It also clarifies the distinction between unconstrained matrix reconstruction and physical RUS inversion: the former searches an entire isospectral orbit, while the latter searches only the elasticity manifold generated by stable Hookean elasticity.

\paragraph*{Relation to RUS developments beyond regular parallelepipeds.}

The formulation above is written for a finite \RR{} representation of a regular sample geometry, but the same inverse spectral viewpoint underlies broader RUS and ultrasonic-resonance developments. For example, RUS has been used for full elastic-constant determination in anisotropic benchmark samples~\cite{Sedlak2014}, extended to less standard experimental contexts and materials~\cite{Ulrich2002ElasticModuli, Li2010HighTempRUS, Geimer2026}, adapted to local ultrasonic resonance spectroscopy for plate inspection~\cite{RusGrosse2020LURS}, and recently generalized to irregularly shaped samples~\cite{Theuss2024}. These developments reinforce the usefulness of treating RUS and related resonance-based ultrasonic methods as spectral inverse problems constrained by material symmetry, stability, and sample geometry.

\section{Supplementary data and benchmark information}
\label{app:supplementary_data_benchmarks}

This appendix summarizes the supplementary information needed to interpret the datasets and benchmarks. Further details on data generation and exploratory analyses are available in Ref.~\cite{CubillosMunoz2025}. The implementation is available in the public repository \href{https://github.com/cubos-d/RUSpectroscopy_Tools}{\texttt{RUSpectroscopy\_Tools}}~\cite{Cubillos2026}, which contains the resources needed to reproduce the forward RUS calculations, data generation, reduced inverse pipeline, and benchmark analyses. Accordingly, this appendix reports only the sampled physical domain, compact spectral diagnostics, and benchmark summaries, while implementation details are delegated to the repository.

\paragraph*{Sampling domain.}
The synthetic datasets were generated by sampling geometries, densities, and elastic constants over prescribed ranges, solving the \RR{} forward problem, and retaining $n_\omega$ positive resonance frequencies. Table~\ref{tab:app_sampling_domain} summarizes the sampled domain. These ranges define the computational domain of the supervised inverse problem and should not be interpreted as a complete classification of possible materials. Stability constraints were imposed on the sampled elastic constants: isotropic samples were generated with positive bulk and shear moduli, while cubic samples were generated in the Born-stable region discussed in Sec.~\ref{subsec:inverse_map}.

\begin{figure}
    \centering
    \includegraphics*[width=0.75\columnwidth]{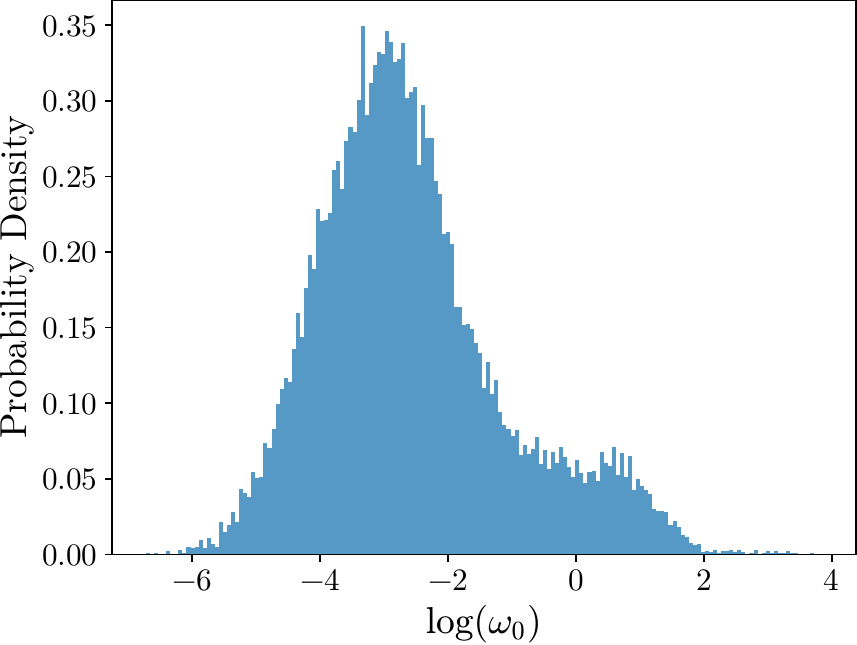}
    \caption{
    Probability distribution of the generated resonance frequencies on a logarithmic scale.
    }
    \label{fig:app_log_frequency_distribution}
\end{figure}

\paragraph*{\RR{} polynomial basis.}
Appendix~\ref{app:rayleigh_ritz_geometry} defines the \RR{} approximation in terms of an abstract finite-dimensional subspace $\Vspace_N\subset H^1(V;\mathbb R^3)$. In the numerical generation of the synthetic spectra, this subspace is specified using polynomial basis functions in normalized coordinates $X=x/L_x$, $Y=y/L_y$, and $Z=z/L_z$. The scalar basis functions are
\begin{equation}
    \phi_{pqr}(X,Y,Z)
    =
    X^pY^qZ^r,
    \label{eq:app_polynomial_basis}
\end{equation}
where $p,q,r\ge 0$, $p+q+r\le N_g$, and $N_g$ is the maximum total polynomial degree retained in the \RR{} expansion. This concrete degree $N_g$ fixes the abstract truncation label $N$ used in $\Vspace_N$. This truncation label should not be confused with $n_\omega$, which denotes the number of measured or selected resonances used in the inversion. The corresponding vector-valued displacement basis is obtained by assigning each scalar basis function to each Cartesian component,
\begin{equation}
    \boldsymbol\Phi_{ipqr}(\vect r)
    =
    \phi_{pqr}(X,Y,Z)\,
    \vect{\hat e}_i,
    \qquad
    i=1,2,3.
    \label{eq:app_vector_polynomial_basis}
\end{equation}
Thus the trial displacement takes the form
\begin{equation}
    \vect u_N(\vect r,t) 
    =
    \sum_{i=1}^{3} \sum_{p+q+r\le N_g} a_{ipqr}(t) \boldsymbol\Phi_{ipqr}(\vect r). 
    \label{eq:app_trial_displacement_xyz_basis}
\end{equation}
Equivalently, the abstract basis index $\alpha$ in App.~\ref{app:rayleigh_ritz_geometry} may be identified with the multi-index $(i,p,q,r)$. The \RR{} coefficient vector $\vect a(t)$ is obtained by collecting all coefficients $a_{ipqr}(t)$ associated with the allowed polynomial triples $(p,q,r)$ and the three displacement components. The number of scalar monomials is
\begin{equation}
    N_b
    =
    \binom{N_g+3}{3}
    =
    \frac{(N_g+1)(N_g+2)(N_g+3)}{6},
\end{equation}
so the full vector-valued \RR{} space has dimension $d_N=3N_b$.
The mass and stiffness matrices used in the generalized eigenvalue problem are the matrix representations of the mass inner product and elastic bilinear form of App.~\ref{app:rayleigh_ritz_geometry} on this polynomial subspace. After solving the generalized eigenvalue problem, the rigid-body zero modes are discarded and the positive eigenvalues are used to construct the reduced spectral variables.

\begin{figure}
    \centering
    \includegraphics*[width=0.9\columnwidth]{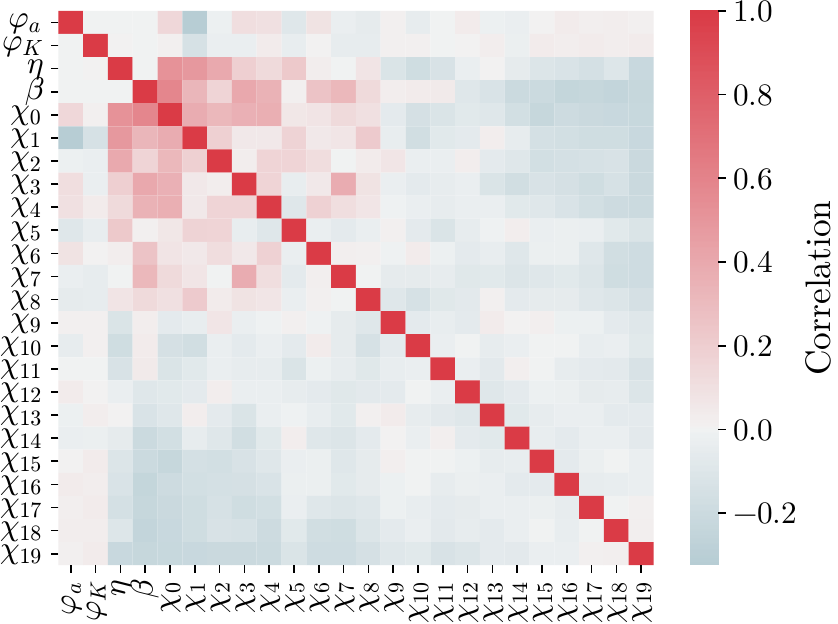}
    \caption{
    Correlation matrix of the reduced cubic features. The variables include the normalized spectral spacings $\{\chi_j\}$ and the aspect-ratio coordinates $(\eta,\beta)$ used as inputs for the cubic angular model. The reduced features show substantially weaker pairwise correlations than the raw spectral representation, indicating that the physics-informed transformation also removes the dominant collinearity in the cubic case.
    }
    \label{fig:app_cubic_reduced_correlation}
\end{figure}

\paragraph*{Spectral diagnostic.}
The main text defines the reduced spectral and geometric variables used by the learning pipeline: the size-normalized spectrum, isotropic ratios $\{\xi_j\}$, cubic normalized spacings $\{\chi_j\}$, and aspect-ratio coordinates $(\eta,\beta)$. Here we include two compact diagnostics from the exploratory analysis. Figure~\ref{fig:app_log_frequency_distribution} shows the log-frequency distribution. Its departure from normality supports the use of transformed spectral variables before regression. Figure~\ref{fig:app_cubic_reduced_correlation} shows the reduced cubic-feature correlation matrix, confirming that the reduced-correlation pattern observed for the isotropic variables in the main text also holds for the cubic spectral representation.

\begin{table}[!b]
    \caption{
    Train, validation, and test metrics for the angular regression models. The isotropic model predicts the scaled angular variable $\phi_K$, and the cubic model predicts $(\phi_{\mathcal K},\phi_a)$.
    }
    \label{tab:app_angular_model_metrics}
    \begin{ruledtabular}
    \begin{tabular}{lllll}
        Angular model & Split & MAE & RMSE & $R^2$ \\
        \hline

        & Train
        & $0.024$
        & $0.038$
        & $0.982$ \\

        Isotropic
        & Validation
        & $0.024$
        & $0.039$
        & $0.981$ \\

        & Test
        & $0.024$
        & $0.040$
        & $0.980$ \\

        \hline

        & Train
        & $0.018$
        & $0.036$
        & $0.973$ \\

        Cubic
        & Validation
        & $0.028$
        & $0.061$
        & $0.924$ \\

        & Test
        & $0.028$
        & $0.062$
        & $0.922$ \\
    \end{tabular}
    \end{ruledtabular}
\end{table}

\paragraph*{Angular-model metrics and learning details.} 
The angular models were trained on the reduced variables defined in the main text. The isotropic dataset contains $252474$ entries with a $60/20/20$ train/validation/test split and inputs $(\eta,\beta,\xi_0,\ldots,\xi_4)$ targeting $\phi_K$. The cubic dataset contains $858387$ entries with a $76/12/12$ split and inputs $(\eta,\beta,\chi_0,\ldots,\chi_{19})$ targeting $(\phi_{\mathcal K},\phi_a)$. The isotropic network uses hidden widths $64,32,8$, while the cubic network uses twelve hidden layers with maximum width $6333$; further implementation details are provided in the public repository. Table~\ref{tab:app_angular_model_metrics} reports the corresponding angular-regression metrics before scale recovery and final elastic-constant reconstruction.


\begin{figure}
    \centering
    \includegraphics*[height=0.38\columnwidth]{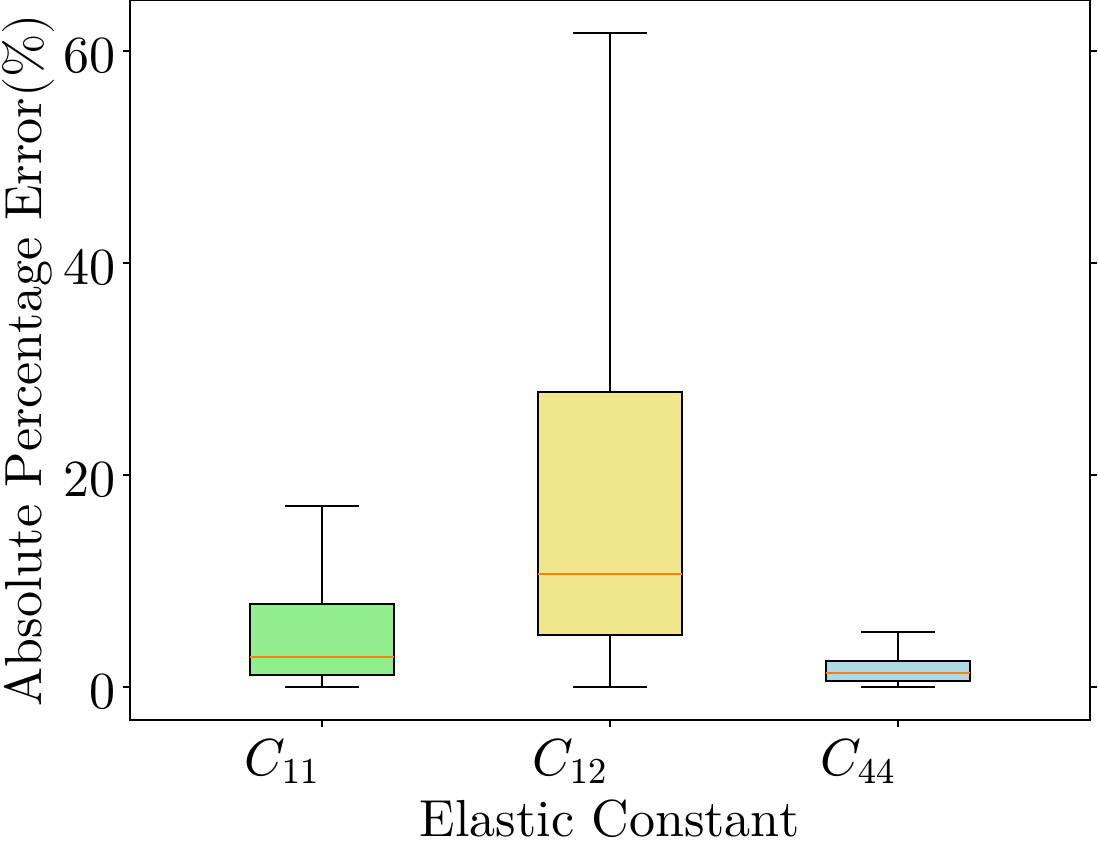}
    \includegraphics*[trim={26pt 0 0 0}, height=0.38\columnwidth]{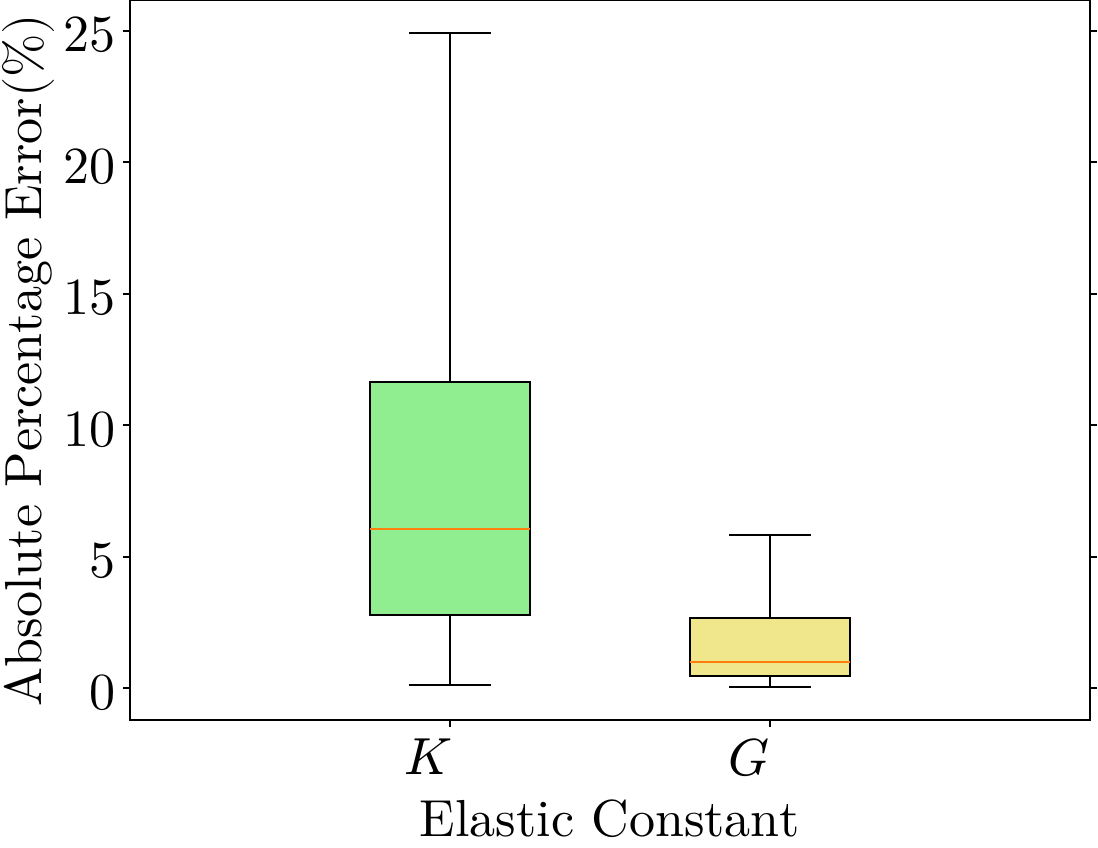}
    \caption{
    Absolute percentage-error distributions for the full cubic and isotropic benchmarks. The left panel shows the cubic reconstruction errors for $C_{11}$, $C_{12}$, and $C_{44}$; the right panel shows the isotropic reconstruction errors for $K$ and $G$. Unlike the fixed-geometry benchmark in Fig.~\ref{fig:absolute_error_distribution}, these distributions retain the aspect-ratio dependence present in the full benchmark.
    }
    \label{fig:app_full_benchmark_ape}
\end{figure}


\paragraph*{Extended benchmark diagnostics.}
The main text reports the full cubic benchmark through parity plots and uses the fixed-geometry benchmark as a compact accuracy diagnostic. For completeness, Fig.~\ref{fig:app_full_benchmark_ape} shows the absolute percentage-error distributions for the full cubic and isotropic benchmarks. These relative-error distributions are useful experimentally, but they require care when the target modulus is small, especially for $C_{12}=\mathcal K-a$.

The larger spread in the full-benchmark percentage-error distributions is consistent with the scope-accuracy tradeoff discussed in the main text. Unlike the fixed-geometry benchmark, the full benchmark retains the dependence on sample dimensions and aspect ratios and samples the thermodynamically stable domain allowed by the physics-informed coordinates. Thus the resulting errors reflect both elastic-ratio reconstruction and geometry-dependent spectral sensitivity. The effect is most visible for $C_{12}$, whose reconstruction involves the cancellation-sensitive combination $C_{12}=\mathcal K-a$.

Because percentage errors can be amplified by small denominators, especially for $C_{12}$, we use absolute-error metrics as the primary numerical summary of the full benchmark. For the full cubic benchmark, the MAE values are $20.37(35.15)~\mathrm{GPa}$, $24.30(41.33)~\mathrm{GPa}$, and $2.13(3.66)~\mathrm{GPa}$ for $C_{11}$, $C_{12}$, and $C_{44}$, respectively. For the full isotropic benchmark, the corresponding values are $6.45(8.42)~\mathrm{GPa}$ and $2.90(5.92)~\mathrm{GPa}$ for $K$ and $G$, respectively. These values show the same qualitative hierarchy as the percentage-error distributions: $C_{44}$ and $G$ are reconstructed most tightly, while $C_{12}$ remains the most sensitive cubic constant.

Material-by-material predictions, percentage errors, and generated benchmark outputs are provided with the public repository~\cite{Cubillos2026}.

\bibliography{rus}

\end{document}